\documentclass[pra,letter,%A4,%twocolumn,%preprint,
showpacs,%preprintnumbers,
floatfix,superscriptaddress]{revtex4}
\usepackage{amsmath}
\usepackage{amsfonts}
\usepackage{amssymb}
\usepackage{graphicx}
\usepackage{epsfig}
\usepackage{dcolumn}
\usepackage{bm}
\usepackage{relsize}
\usepackage[usenames,dvipsnames]{color}	 % For coloring the text and make available the dvips named colors. See the file dvips_std_colors.pdf for more.
\usepackage[colorlinks=true,breaklinks=true,bookmarksopen=true,bookmarksnumbered=true,citecolor=blue,linkcolor=red,hyperfootnotes=false]{hyperref}  % Change the colors to any dvips named color you wish
\usepackage{soul}
\begin{document}
\title{Spin asymmetry for the elastic scattering of polarized electrons from Zn, Cd, and Hg}
\author{Mehrdad Adibzadeh}
\email{madibzadeh@uwf.edu}
\affiliation{Department of Physics, University of West Florida, Pensacola, FL 32514, USA}
\author{Constantine E. Theodosiou}
\email[Corresponding author: ]{ constant.theodosiou@manhattan.edu}
\affiliation{Department of Mathematics and Physics, Manhattan University, Riverdale, NY 10471, USA.}
\date{\today}

\begin{abstract}
We present an extensive set of theoretical results for spin asymmetry in the form of Sherman functions for the elastic scattering of electrons by zinc, cadmium, and mercury. This study extends the application of our earlier method of calculations, which we previously employed for stable inert gases and alkaline-earth-metals to these three atoms. Our predictions are in adequate agreement with experimental values and precise theoretical results.
\end{abstract}
\maketitle

\section{Introduction}
The investigation of spin asymmetry in elastic electron scattering from atoms underscores the extent of spin-dependent interactions in electron-atom collisions. In the case of closed-shell atoms, which represent spin-zero targets, the spin asymmetry effects are due to the spin-orbit interactions. Thus the theoretical studies of spin asymmetry in elastic scattering from heavy closed shell atoms such as zinc, cadmium and mercury signifies the subtleties of relativistic effects and the antisymmetrization of the wave function.

Among the three atoms, mercury is the most studied through atomic collisions. Zinc and cadmium, although not as favorite as mercury, have been experimentally studied from the onset of the investigations into atomic collisions. The early studies, however, were all concentrated on cross sections. The studies into spin polarization effects followed later on and the interest in both experimental and theoretical approaches made an appearance.

For zinc, there is only one experimental work on the study of spin polarization of elastically scattered electrons, that of Bartsch et al.~\cite{Bartsch1992} for energies up to 14 eV. The theoretical investigations on the subject of spin polarization of elastically scattered electrons off zinc atom are those of McEachran and Stauﬀer~\cite{McEachran1992}, Kumar et al.~\cite{Kumar1994}, Szmytkowski and Sienkiewicz~\cite{Szmytkowski1994} and Bostock et al.~\cite{Bostock2012}.

In the case of cadmium, the only experimental work is that of Bartsch et al.~\cite{Bartsch1992} for energies up to 9 eV. The body of theoretical elastic spin asymmetry calculations on cadmium includes the works of Nahar~\cite{Nahar1991}, McEachran and Stauﬀer~\cite{McEachran1992}, Kumar et al.~\cite{Kumar1994}, Szmytkowski and Sienkiewicz~\cite{Szmytkowski1994},  Berrington et al.~\cite{Berrington2012} and Haque et al.~\cite{Haque2021}.

Mercury, on the other hand, has enjoyed more attention from experiment and theory. There exist several  spin polarization experimental investigations of elastically scattered electrons from mercury. They include works of Deichsel~\cite{Deichsel1961}, Jost and Kessler~\cite{Jost1966}, Deichsel et al.~\cite{Deichsel1966},  Eitel et al.~\cite{Eitel1967}, Eitel and Kessler~\cite{Eitel1971}, Hanne et al.~\cite{Hanne1972}, D\"{u}weke et al.~\cite{Duweke1976}, Albert et al.~\cite{Albert1977}, Hanne et al.~\cite{Hanne1980}, Berger et al.~\cite{Berger1981}, M\"{o}llenkamp et al.~\cite{Mollenkamp1984}, Berger and Kessler~\cite{Berger1986}, Kaussen et al.~\cite{Kaussen1987} and D\"{u}mmler et al.~\cite{Dummler1992}.The theoretical calculations of the subject include those of Bunyan~\cite{Bunyan1963}, Bunyan and Schonfelder~\cite{Bunyan1965}, Fink and Yates~\cite{Fink1969}, Bartschat et al.~\cite{Bartschat1984}, Haberland and Fritsche~\cite{Haberland1987}, McEachran and Stauffer~\cite{McEachran1987}, Sienkiewicz~\cite{Sienkiewicz1990}, Fritsche et al.~\cite{Fritsche1992}, Szmytkowski and Sienkiewicz~\cite{Szmytkowski1994}, Kelemen and Remeta~\cite{Kelemen2012}, and Haque et al.~\cite{Haque2017,Haque2021}

The present work intends to test our semi empirical approach (already applied to closed shell inert gas atoms and quasi-two-electron stable alkaline earth metal atoms~\cite{Adibzadeh2005, Adibzadeh2004,Adibzadeh2024}) for zinc, cadmium and mercury. In doing so, we will compare our results with only a reasonable number of dependable theoretical works, in addition to all available experimental data, in order to keep graphs readable.

\section{Brief Review of the Theoretical and Computational Approach}
In this work, we apply the same method of calculations as that described in our previous papers~\cite{Adibzadeh2004, Adibzadeh2005, Adibzadeh2024}, i.e. the standard method of partial-wave expansion in potential scattering. We obtained the phase shifts by solving the stationary Dirac equation, for which the choices of central static atomic, exchange, and polarization potentials were determined through an extensive analysis. The present choices for those potentials turned out to be the same as those of Refs.~\cite{Adibzadeh2004, Adibzadeh2024}. 

To determine the combination for potentials, we performed an exhaustive analysis and comparison with the collection of all appropriate experimental and theoretical results on elastic electron scattering by inert gases and alkaline-earth metals. As a result we were able to recommend a combination of the central static atomic, exchange, and polarization potentials, which produced a consistent agreement between the calculated cross sections and reliable experimental and theoretical results at different projectile energies.  

In the present work, we applied the same diligence to Zn, Cd and Hg to confirm our earlier choices for the only free parameter in our calculations: the cutoff radius of the polarization potential. Our analysis showed that the most consistent combination of potentials for Zn, Cd, and Hg was to be the Dirac-Slater atomic potential, the semi-classical exchange potential expression by Furness and McCarthy~\cite{Furness1973} and the Buckingham-type II polarization potential in the form
\begin{equation}
V_P(r) =  - \frac{\alpha _d}{2(r^2 + r_{c}^2 )^2}\, .
\end{equation}
In the above equation, $ \alpha _d $ and $r_c $ are the static atomic polarizability and the cutoff radius, respectively. The values for static polarizability, $ \alpha _d $, were taken from the theoretical work of Kolb et al.~\cite{Kolb1982}. 

Similar to our previous works, we required $r_c$ to be a smooth, continuous and finite function of the scattered electron energy $E$. We further confirmed the functional behavior of this energy-dependent cutoff radius to be similar to those we used for alkaline-earth-metal atoms ~\cite{Adibzadeh2004,Adibzadeh2024} through comparisons with dependable theoretical and experimental data. Our careful analysis determined the value for the cutoff radius for zinc to be (in atomic units)

\begin{equation}
r_c (E) = \left\{ {\begin{array}{*{20}c}
   {\frac{1}{3}\ln (\frac{E}{\mathcal{R}}) + \left\langle r \right\rangle _{4s} }
   & {E \ge 45{\rm{ eV}}},  \\\\
   {{\rm{3}}{\rm{.85}}} & {E < 45{\rm{ eV}}},  \\
\end{array}} \right.
\end{equation}

for cadmium,

\begin{equation}
r_c (E) = \left\{ {\begin{array}{*{20}c}
   {\frac{1}{3}\ln (\frac{E}{\mathcal{R}}) + \left\langle r \right\rangle _{5s} }
   & {E \ge 30{\rm{ eV}}},  \\\\
   {{\rm{4}}{\rm{.0}}} & {E < 30{\rm{ eV}}},  \\
\end{array}} \right.
\end{equation}

and for mercury,

\begin{equation}
r_c (E) = \left\{ {\begin{array}{*{20}c}
   {\frac{1}{3}\ln (\frac{E}{\mathcal{R}}) + \left\langle r \right\rangle _{6s} }
   & {E \ge 20{\rm{ eV}}},  \\\\
   {{\rm{3}}{\rm{.6}}} & {E < 20{\rm{ eV}}}.  \\
\end{array}} \right.
\end{equation}

Here $E$ is the energy of the incident electron in eV, $\mathcal{R}$ is the Rydberg 
constant ($\mathcal{R}$ = 13.605 691 72 eV), and $\left\langle r \right\rangle
_{4s} = \text{2.680 }a_0$,  $\left\langle r \right\rangle
_{5s} = \text{2.978 }a_0 $ and  $\left\langle r \right\rangle
_{6s} = \text{3.045 }a_0 $ are the expectation values of zinc's $4s$ shell, cadmium's $5s$ shell, and mercury's $6s$ shell radii, respectively. The cutoff radii for low energies are set to constant values to avoid the logarithmic anomaly. The reader may find more details on this constant value and its behavior below an energy threshold in Ref.~\cite{Adibzadeh2004}. To determine the low energy constant values for $r_c$, we made numerous comparisons with accurate theoretical data for integrated cross section as guidance.

Once the Dirac Hamiltonian is determined and the Dirac equation is solved, we can obtain the spin-up $\delta _l^+ $ and spin-down $\delta _l^- $ phase shifts, which are used to determine the direct, $f$, and spin-flip, $g$, scattering amplitudes

\begin{equation}
f(\theta ,k) =  \frac{1}
{k}\sum\limits_{l = 0}^\infty  {\left[ {(l + 1)e^{i\delta _l^ +  (k)} \sin \delta _l^ +  (k)} \right.} 
\left. { + le^{i\delta _l^ -  (k)} \sin \delta _l^ -  (k)} \right] \\
P_l (\cos \theta ) \, ,
\end{equation}

and

\begin{equation}
g(\theta ,k) = \frac{1}
{k}\sum\limits_{l = 0}^\infty  {\left[ {e^{i\delta _l^ -  (k)} \sin \delta _l^ -  (k)} \right.}
  \left. { - e^{i\delta _l^ +  (k)} \sin \delta _l^ +  (k)} \right]P_l^1 (\cos \theta )\, . 
\end{equation}

In the above equations, $k$ and $\theta$ are the wave number of the incident electron and the scattering angle, respectively. The Sherman function (SF), which describes the measured spin asymmetries in the number of scattered electrons is expressed in terms of scattering amplitudes as \cite{Sherman1956,Kessler1969}
\begin{equation}
S (\theta ,k) = i\frac{{f(\theta ,k)g^* (\theta ,k)%
 - f^*(\theta,k)g(\theta ,k)}}{{\left| {f(\theta ,k)} \right|^2%
 + \left| {g(\theta ,k)} \right|^2 }}.
\end{equation}

Once again, we used our modified version of the Dirac-Slater code by Salvat et al.~\cite{Salvat1995} in our calculations and throughout this work and for all considered energies, we used up to 150 partial-wave phase shifts.

\section{Results and Discussion}
Since we would like to compare our results with all available experimental measurements, to maintain graph readability, we will limit the comparisons to most recent and reliable data whenever possible. We also avoid the placement of error bars on experimental data if the uncertainty may be encompassed through an appropriate size of the marker.

\subsection{Zinc}
\begin{figure*}
\includegraphics{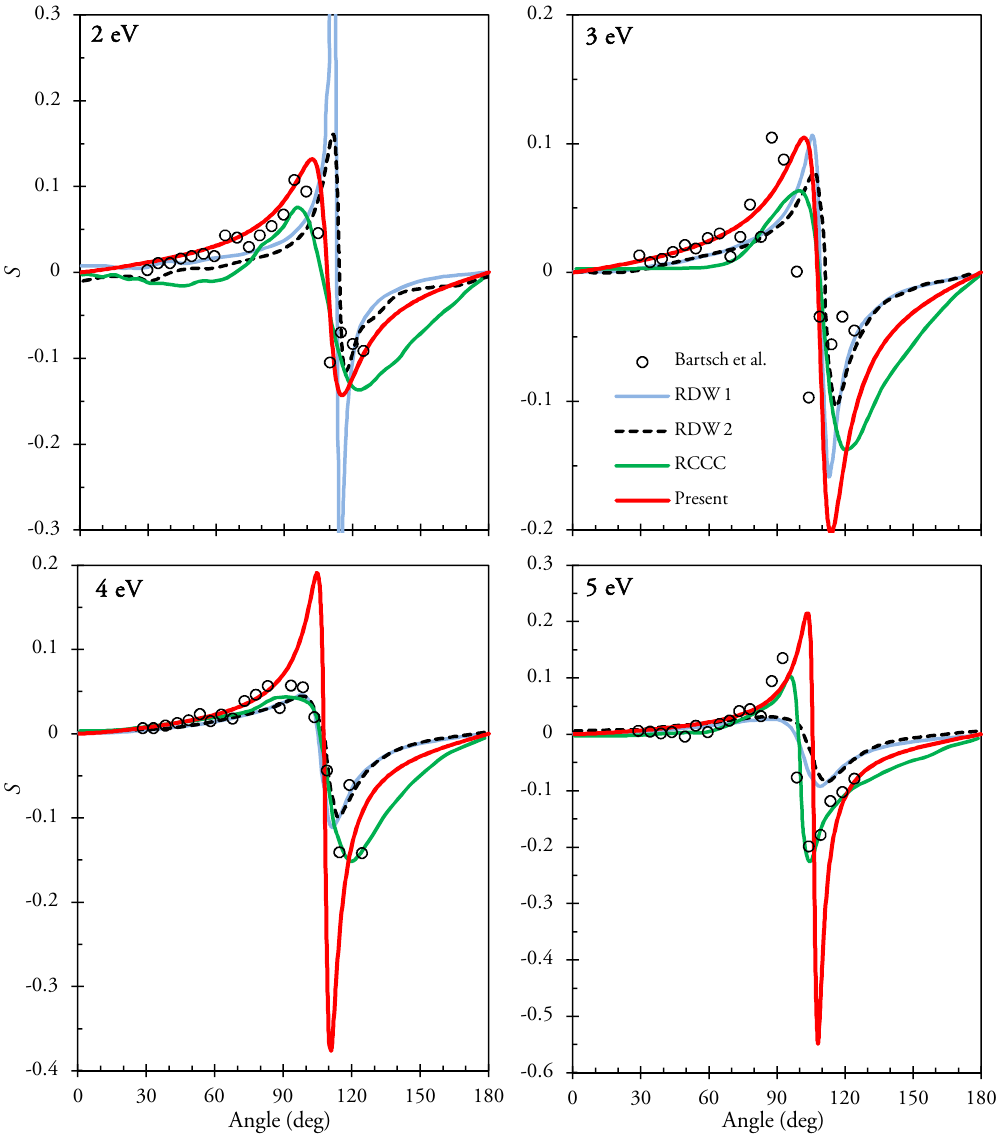}
\caption{\label{fig:zn1} Sherman function for electron scattering from zinc at 2, 3, 4 and 5 eV: The legend in the figure describes markers for Present work; Experiment: Bartsch et al.~\cite{Bartsch1992}; Other theoretical: 66-state RCCC~\cite{Bostock2012}, RDW 1~\cite{McEachran1992} and RDW 2~\cite{Szmytkowski1994}.}
\end{figure*}

\begin{figure*}
\includegraphics{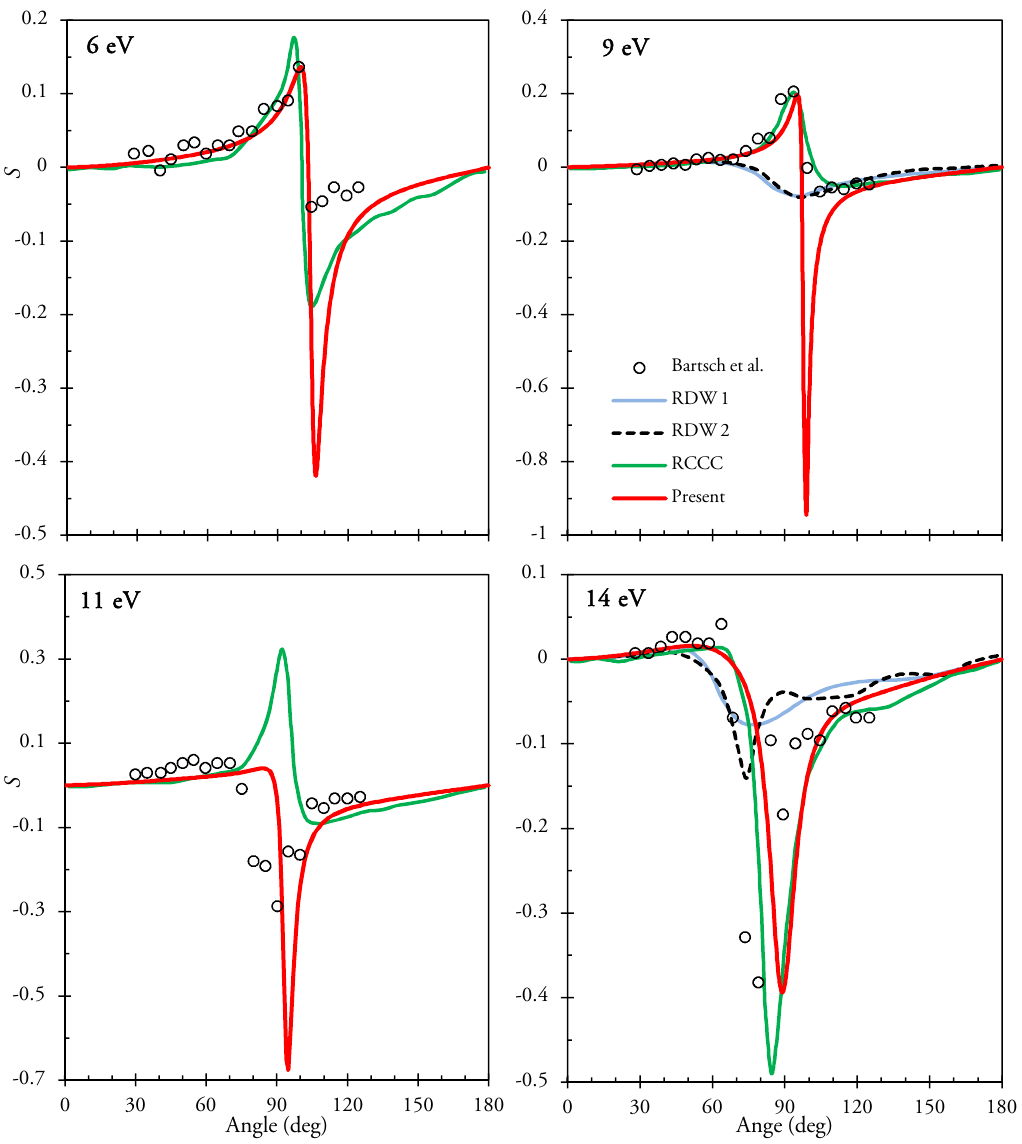}
\caption{\label{fig:zn2} Same as for figure~\ref{fig:zn1} but at 6, 9, 11 and 14 eV projectile energies.}
\end{figure*}

\begin{figure*}
\includegraphics{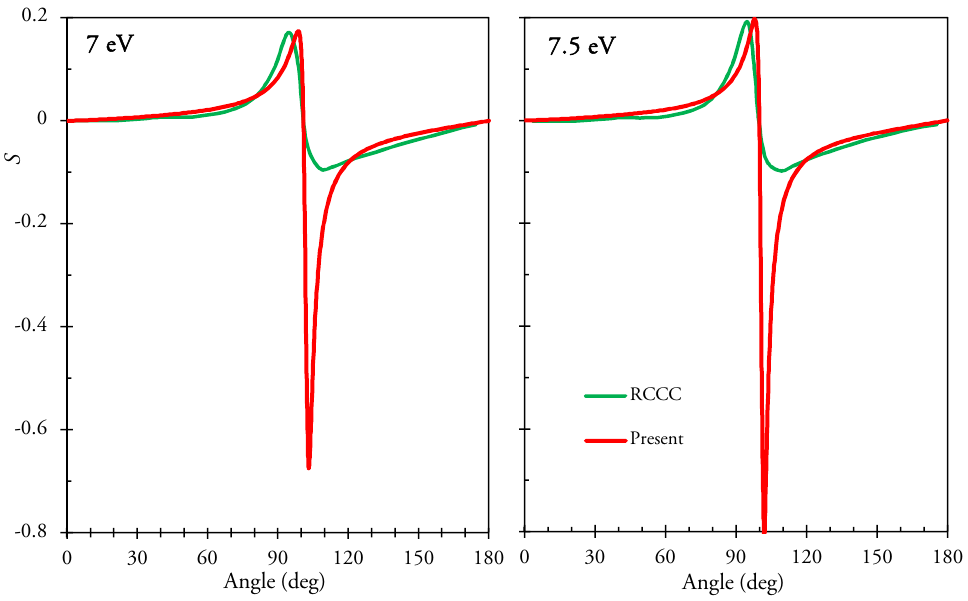}
\caption{\label{fig:zn3} Comparisons between present Sherman function values at 7 and 7.5 eV with those of 66-state RCCC~\cite{Bostock2012} calculations.}
\end{figure*}

The experimental data on spin asymmetry for elastic electron scattering off zinc are limited to those of Bartcsh et al.~\cite{Bartsch1992} for a few energies between 2 eV and 14 eV. At impact energies for which experimental data are available, we compare our Sherman functions with the experiment and other theoretical results in figures~\ref{fig:zn1} and \ref{fig:zn2}. These comparisons include the relativistic distorted-wave (RDW) calculations of McEachran and Stauffer~\cite{McEachran1992} and Szmytkowski and Sienkiewicz~\cite{Szmytkowski1994} and the 66-state relativistic convergent close coupling (RCCC) formulation of Bostock et al.~\cite{Bostock2012}. In figure~\ref{fig:zn3}, we exclusively compare with the available RCCC calculations at 7 and 7.5 eV, for which no experimental data are available.

The present Sherman function values at 2 eV impact energy are in outstanding agreement with experiment, as seen in figure~\ref{fig:zn1}. This agreement is to a lesser extent on display at 3 eV as well. Nonetheless, the agreement of our Sherman functions at 4 and 5 eV with experiment is mostly limited to the forward direction, while the RCCC calculations present excellent agreement with experiment.
For impact energies 6 and 9 eV, on figure~\ref{fig:zn2}, our Sherman functions again show excellent agreement with experiment and the RCCC results in the forward direction. This is while both RDW calculations at 9 eV indicate a considerable disagreement with experiment.

The most interesting comparisons, however, are those at 11 and 14 eV impact energies in figure~\ref{fig:zn2}. At 11 eV, our values show a better agreement with experiment than those of the RCCC calculations. As touched upon by Ref.~\cite{Bostock2012}, the discrepancy between the RCCC Sherman function and experiment at 11 eV, is due to $3d^{10}$ excitations which were not modeled accurately by the RCCC formalism. Considering our single-channel method of calculations does not include the effects of excited states on elastic spin asymmetry, our agreement with experiment is quite interesting. At 14 eV, the present Sherman functions display a comparable agreement with experiment as that between RCCC and experiment.
In figure~\ref{fig:zn3}, we compare our Sherman functions at 7 and 7.5 eV with those of the RCCC calculations with discrepancies to note. These slight differences are the positions of the maxima and the values of the minima of the Sherman function at the two energies, which were observed also at other impact energies.

\begin{figure*}
\includegraphics{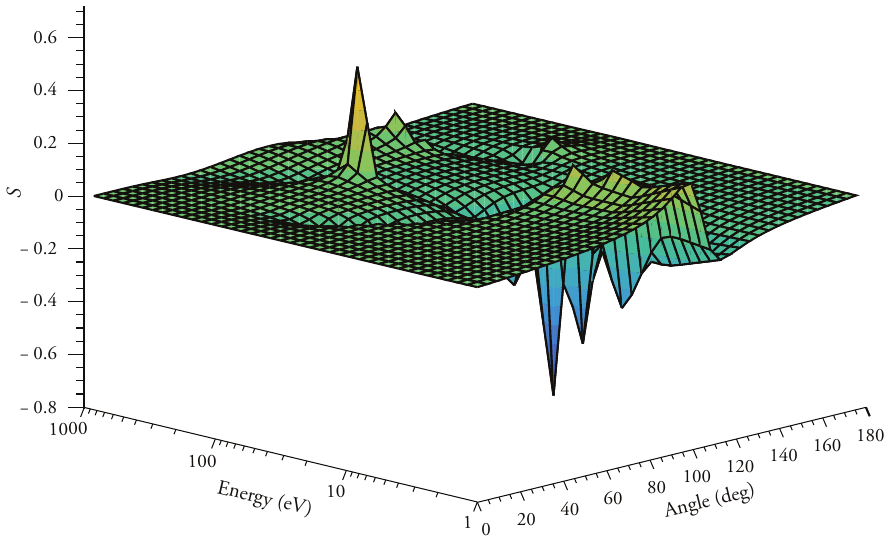}
\caption{\label{fig:zn4} A three-dimensional view of Sherman function for electron scattering from zinc.}
\end{figure*}

Finally, a three-dimensional (3D) graph of present Sherman function versus impact energy and scattering angle for elastic $e$-Zn scattering is provided in figure~\ref{fig:zn4} to obtain a comprehensive view of the behavior of $S(\theta,E)$.

\subsection{Cadmium}

\begin{figure*}
\includegraphics{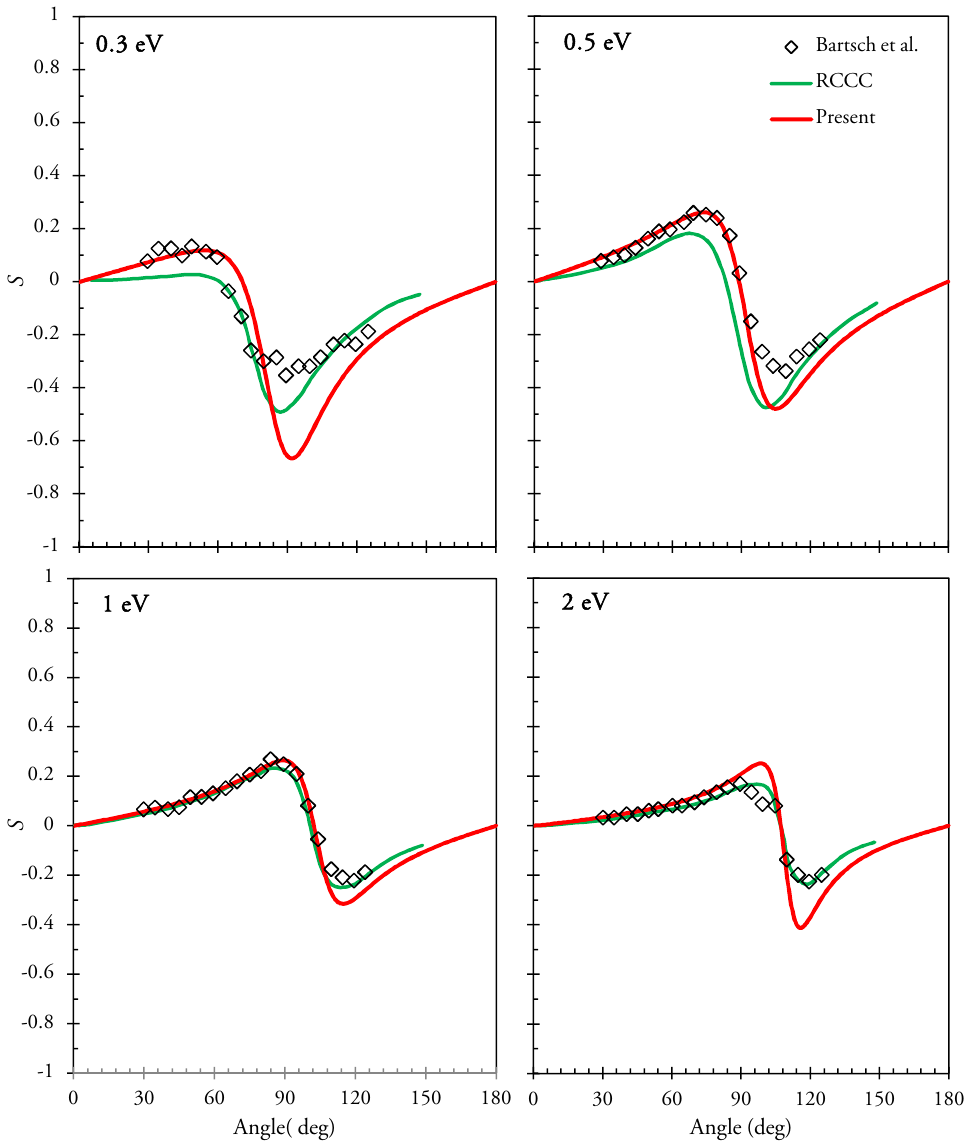}
\caption{\label{fig:cd1} Sherman function for electron scattering from cadmium at 0.3, 0.5, 1 and 2 eV: The legend in the figure describes markers for Present work; Experiment: Bartsch et al.~\cite{Bartsch1992}; Other theoretical: 55-state RCCC~\cite{Berrington2012} calculations.}
\end{figure*}

\begin{figure*}
\includegraphics{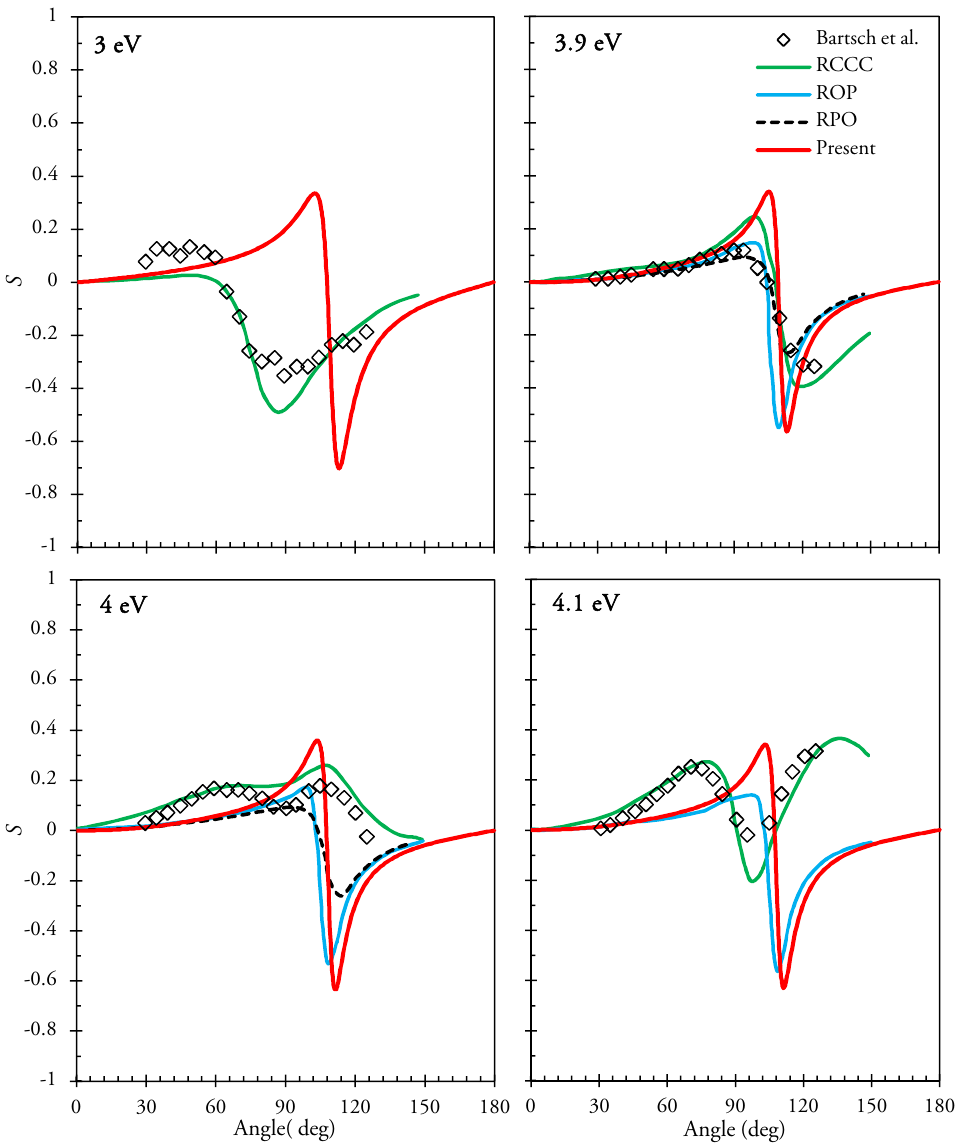}
\caption{\label{fig:cd2} Same as for figure~\ref{fig:cd1} but at at 3, 3.9, 4 and 4.1 eV projectile energies. Additional theoretical works are ROP~\cite{Berrington2012} and RPO~\cite{Szmytkowski1994} calculations.}
\end{figure*}

\begin{figure*}
\includegraphics{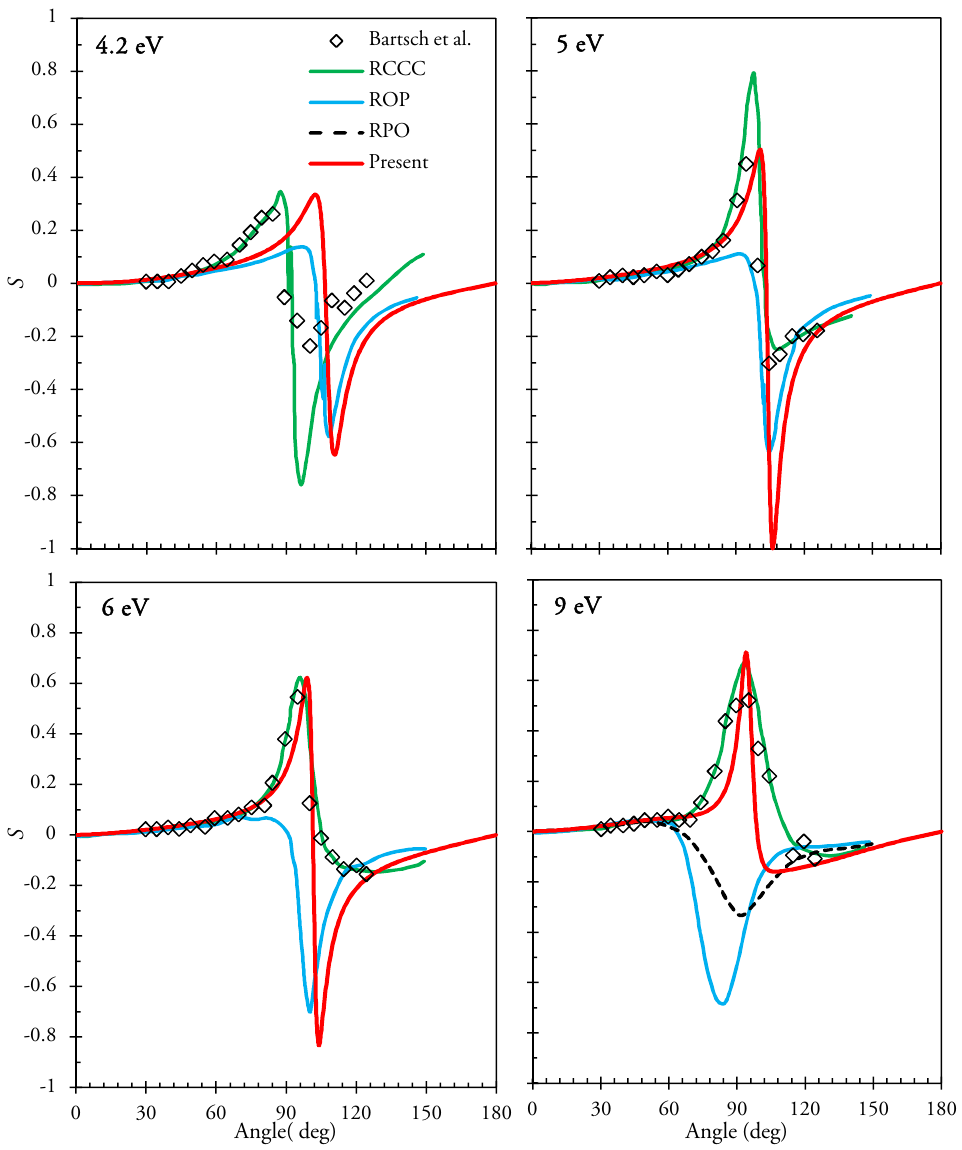}
\caption{\label{fig:cd3} Same as for figure~\ref{fig:cd2} but at 4.2, 5, 6 and 9 eV projectile energies.}
\end{figure*}

The measurements of Bartcsh et al.~\cite{Bartsch1992} are the only experimental data on spin asymmetry for elastic electron scattering off cadmium. We compare our Sherman functions, in figures~\ref{fig:cd1}, \ref{fig:cd2} and \ref{fig:cd3}, with all experimental data in conjunction with other theoretical works, which include the 55-state RCCC method and the relativistic optical potential (ROP) calculations of Berrington et al.~\cite{Berrington2012} and the relativistic polarized orbital approach of Szmytkowski and Sienkiewicz~\cite{Szmytkowski1994}. 
Very good agreement with experiment is observed for energies up to 2 eV, in particular in the forward direction. The agreement at these energies with RCCC results is at least qualitative if not very good. For energies above 2 eV, in particular approaching the $(5s5p)$ $^3 P_{0,1,2}$ excited states, and the $(5s5p^2)$ $^3D_{3/2,5/2}$ negative ion state, our agreement with experiment is rather spotty% if it exists at all
. Setting aside the accurate RCCC predictions, other single-channel calculations do not agree with experiment at these energies either.

\begin{figure*}
\includegraphics{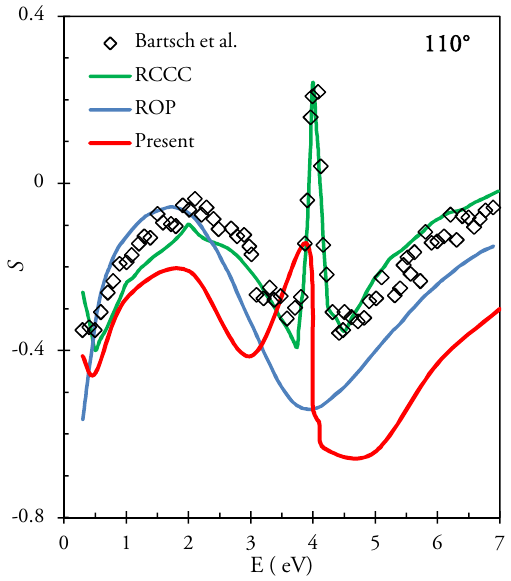}
\caption{\label{fig:cd4} Sherman function for electron scattering from cadmium at scattering angle $110^\circ$ versus impact energy. The legend in the figure describes markers for Present work; Experiment: Bartsch et al.~\cite{Bartsch1992}; Other theoretical: 55-state RCCC~\cite{Berrington2012} and ROP~\cite{Berrington2012} calculations.}
\end{figure*}

Our Sherman functions at 5 and 6 eV impact energies, show an excellent agreement with experiment in the forward direction, while disagreeing by a few degrees on the location of the maximum with RCCC Sherman functions. At both energies, ROP results show considerable discrepancies with experiment. Our Sherman function at 9 eV displays a remarkable agreement with experiment while RCCC’s is the only other calculation in agreement with experiment at this energy.
In figure~\ref{fig:cd4}, we present our Sherman function values at $110^\circ$ as a function of energy in comparison with the experiment and the theoretical results from the RCCC and ROP approaches. The swift rise in Sherman function values at $110^\circ$ between 3.7 and 4.2 eV is due to the formation of negative ion resonances and excited states in this energy interval (discussed in Ref.~\cite{Bostock2011}). Our Sherman function at $110^\circ$, shows a displaced ascend while following the experiment and RCCC through a downward bias.  The overall energy-behavior of our results is good in spite of the single-channel nature of the approach.

\begin{figure*}
\includegraphics{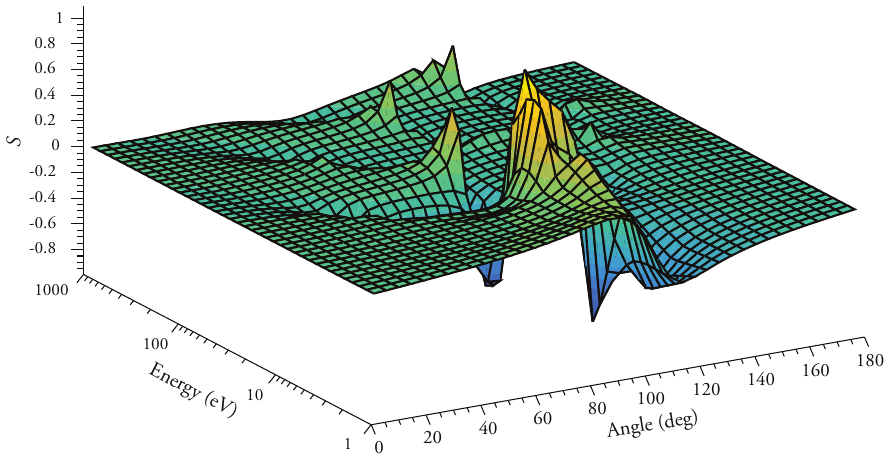}
\caption{\label{fig:cd5} A three-dimensional view of Sherman function for electron scattering from cadmium.}
\end{figure*}

The 3D graph of Sherman function against impact energy and scattering angle for elastic $e$-Cd scattering is presented in figure~\ref{fig:cd5}.

\subsection{Mercury}
As mentioned in the introduction for mercury there is a plethora of spin asymmetry measurements, some of them going back to 1956. The work of D\"{u}weke et al.~\cite{Duweke1976} is limited to energies below 4 eV and that of D\"{u}mmler et al.~\cite{Dummler1992} to less than 2 eV.  Above 6 eV impact energy, we have the works of Kaussen et al.~\cite{Kaussen1987} (6-24, and 180 eV), Hanne et al.~\cite{Hanne1980} (8, 10, 12, and 18 eV), and Deichsel et al.~\cite{Deichsel1966} (3.5, 7, 23, and 45 eV). At intermediate energies, there exist the works of Eitel et al.~\cite{Eitel1967} (18.3, 25, 30, 50, and 180 eV) and that of Berger and Kessler~\cite{Berger1986} (25, 35, 50, and 150 eV). Finally, at the upper range of impact energies we have the works of Jost and Kessler~\cite{Jost1966} (180 to 340 eV) and Eitel et al.~\cite{Eitel1967} (100 - 2000 eV).  
Thus we have experimental data in a wide range of impact energies to compare our theoretical predictions with.

\begin{figure*}
\includegraphics{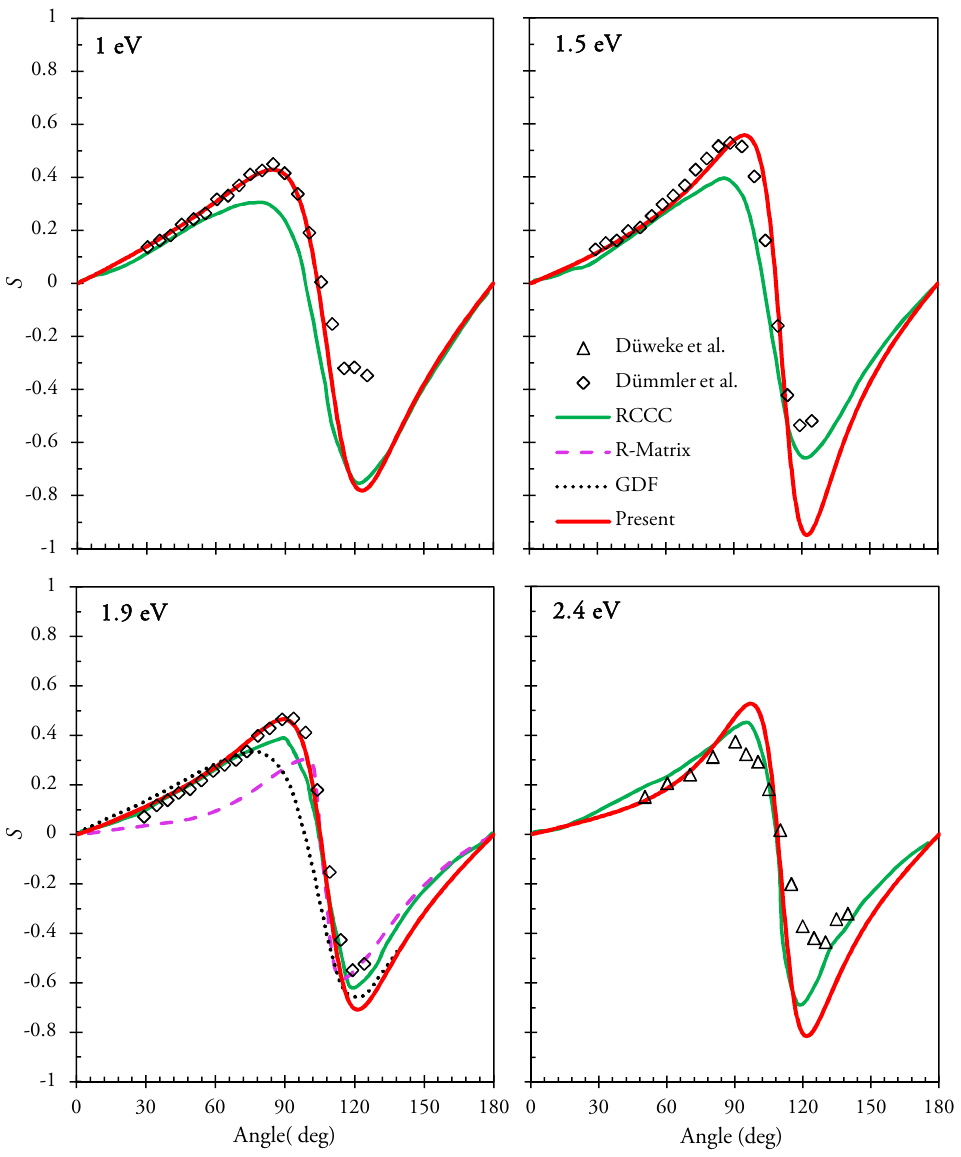}
\caption{\label{fig:hg1} Sherman function for electron scattering from mercury at 1, 1.5, 1.9 and 2.4 eV: The legend in the figure describes markers for Present work; Experiment: D\"{u}weke et al.~\cite{Duweke1976} and D\"{u}mmler et al.~\cite{Dummler1992}; Other theoretical: 66-state RCCC~\cite{Bostock2012}, R-matrix~\cite{Bartschat1984} and GDF~\cite{Haberland1987}.}
\end{figure*}

\begin{figure*}
\includegraphics{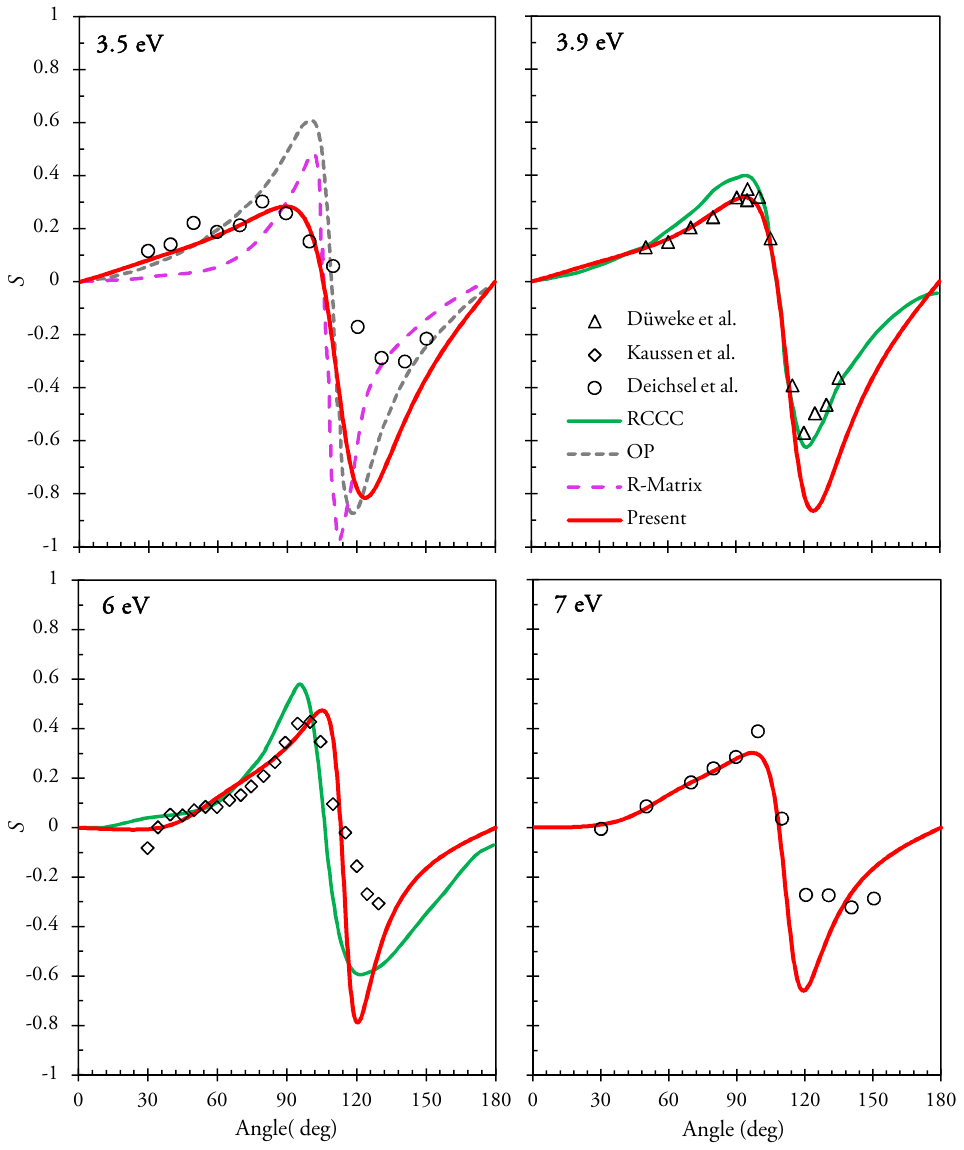}
\caption{\label{fig:hg2} Same as for figure~\ref{fig:hg1} but at 3.5, 3.9, 6 and 7 eV projectile energies. Additional experiments: Kaussen et al.~\cite{Kaussen1987} and Deichsel et al.~\cite{Deichsel1966}; Additional theoretical: OP~\cite{Haque2017}.}
\end{figure*}

\begin{figure*}
\includegraphics{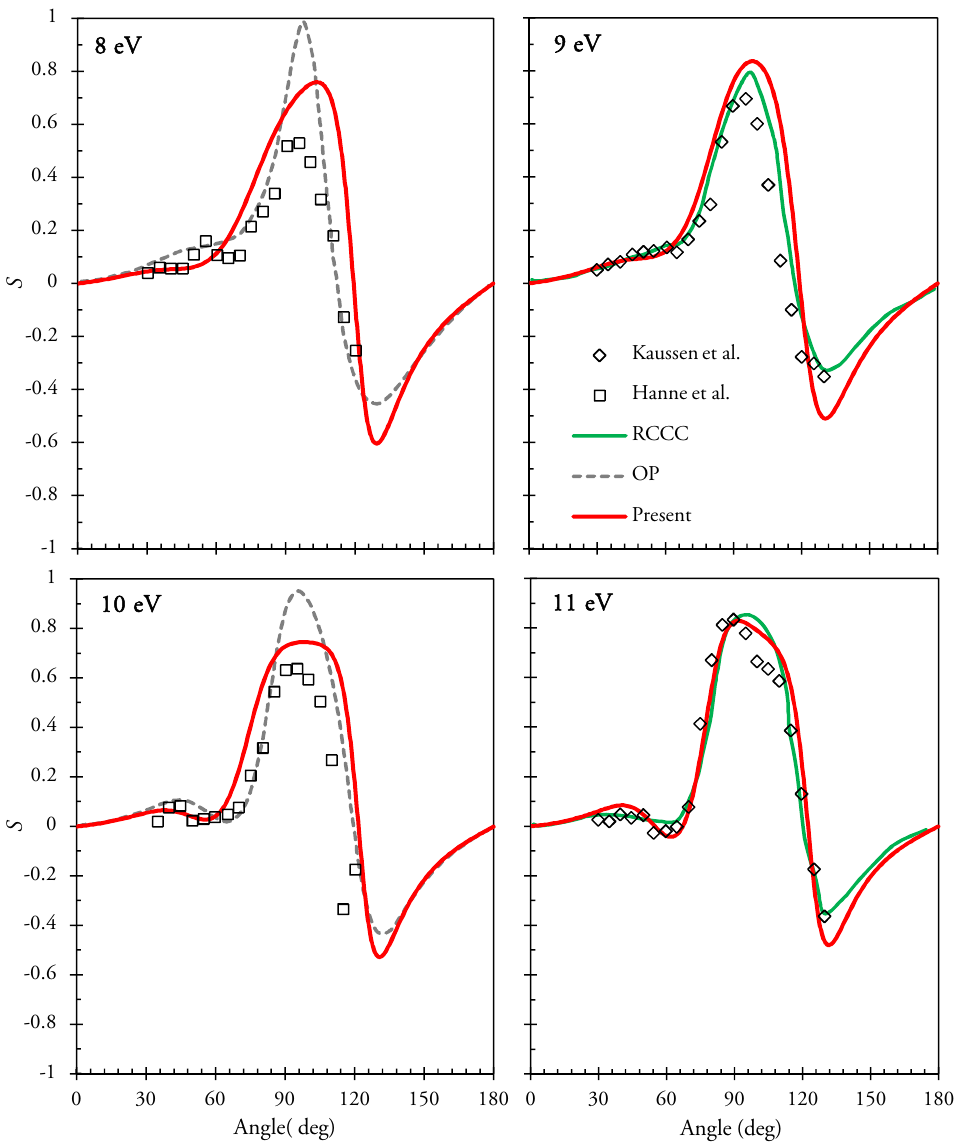}
\caption{\label{fig:hg3} Same as for figure~\ref{fig:hg2} but at 8, 9, 10 and 11 eV projectile energies. Additional experiment is Hanne et al.~\cite{Hanne1980}.}
\end{figure*}

\begin{figure*}
\includegraphics{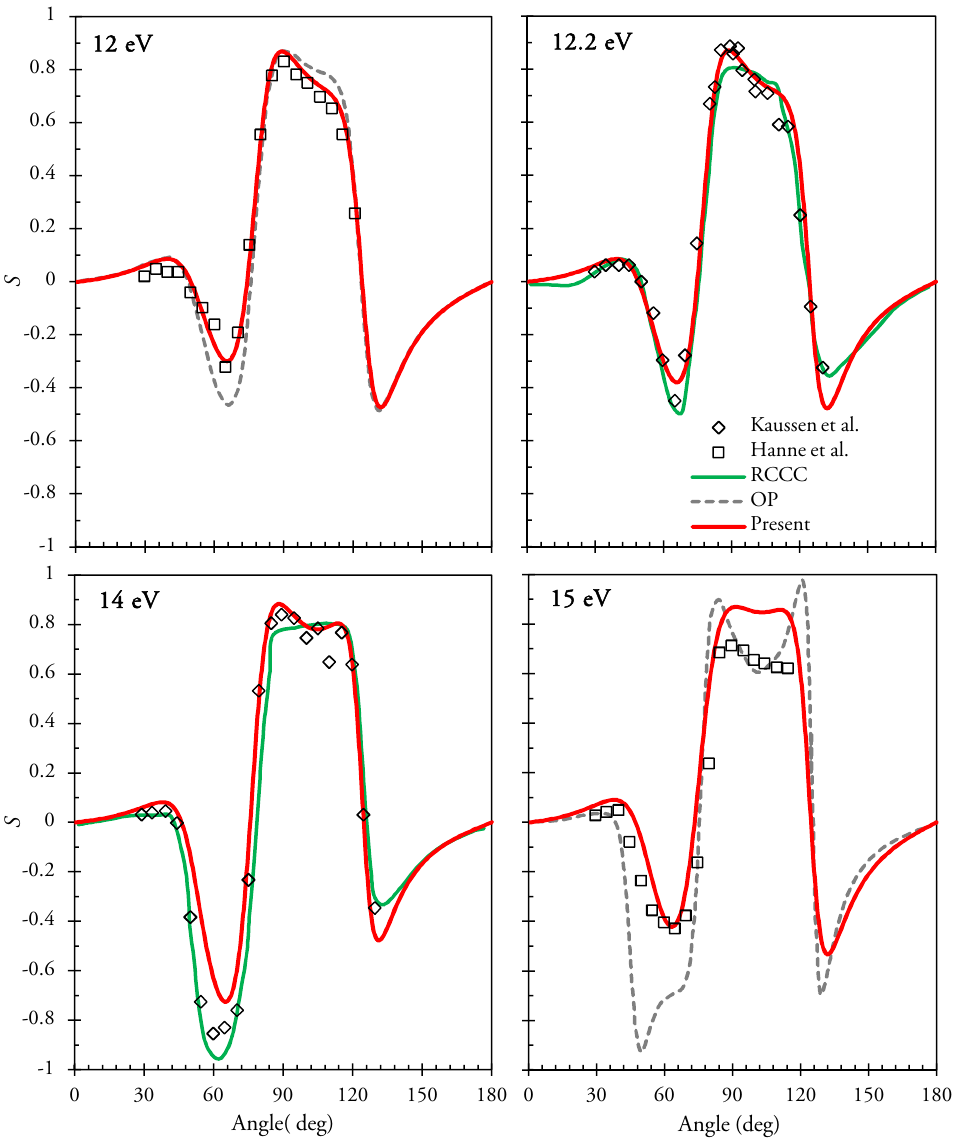}
\caption{\label{fig:hg4} Same as for figure~\ref{fig:hg3} but at 12, 12.2, 14 and 15 eV projectile energies.}
\end{figure*}

\begin{figure*}
\includegraphics{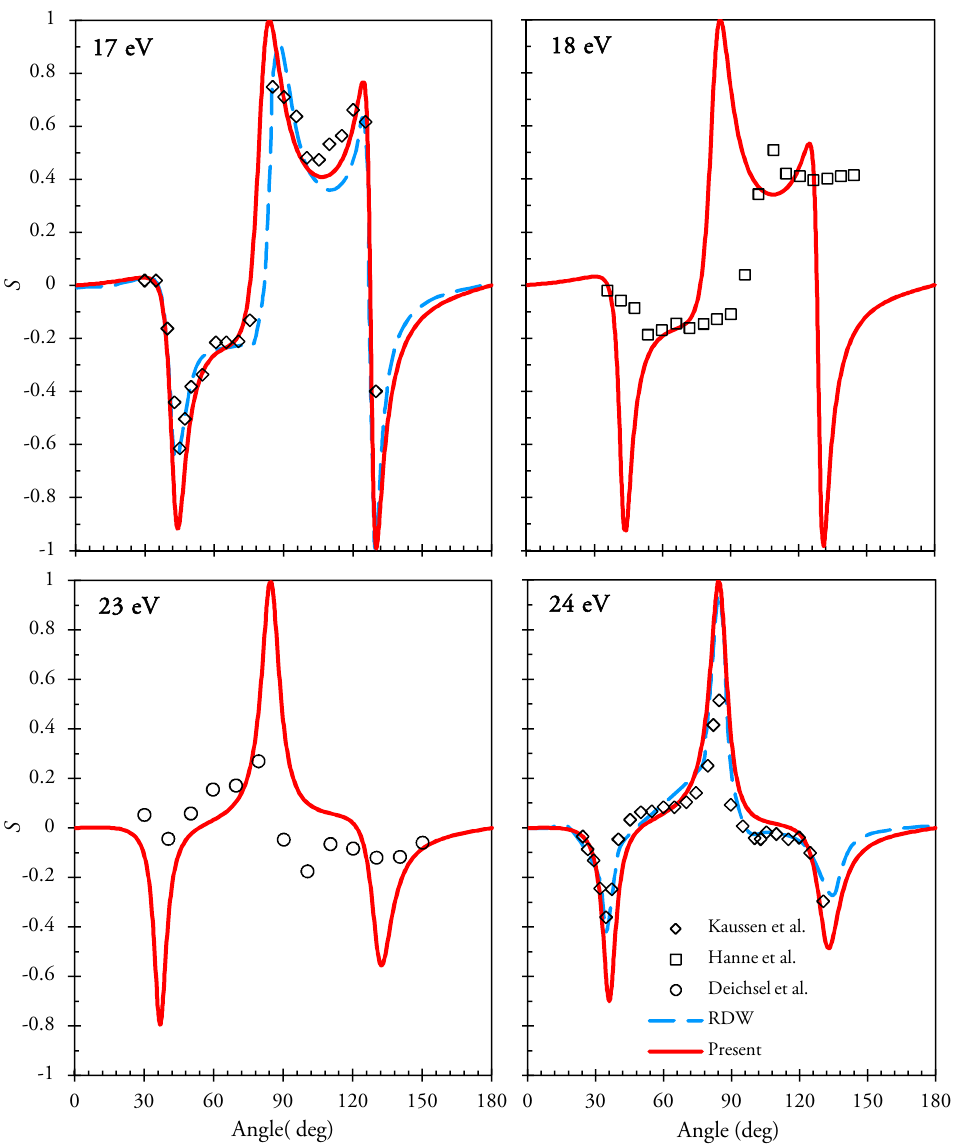}
\caption{\label{fig:hg5} Same as for figures~\ref{fig:hg2} and \ref{fig:hg4} but at 17, 18, 23 and 24 eV projectile energies. Additional theoretical work is RDW~\cite{McEachran1987}.}
\end{figure*}

\begin{figure*}
\includegraphics{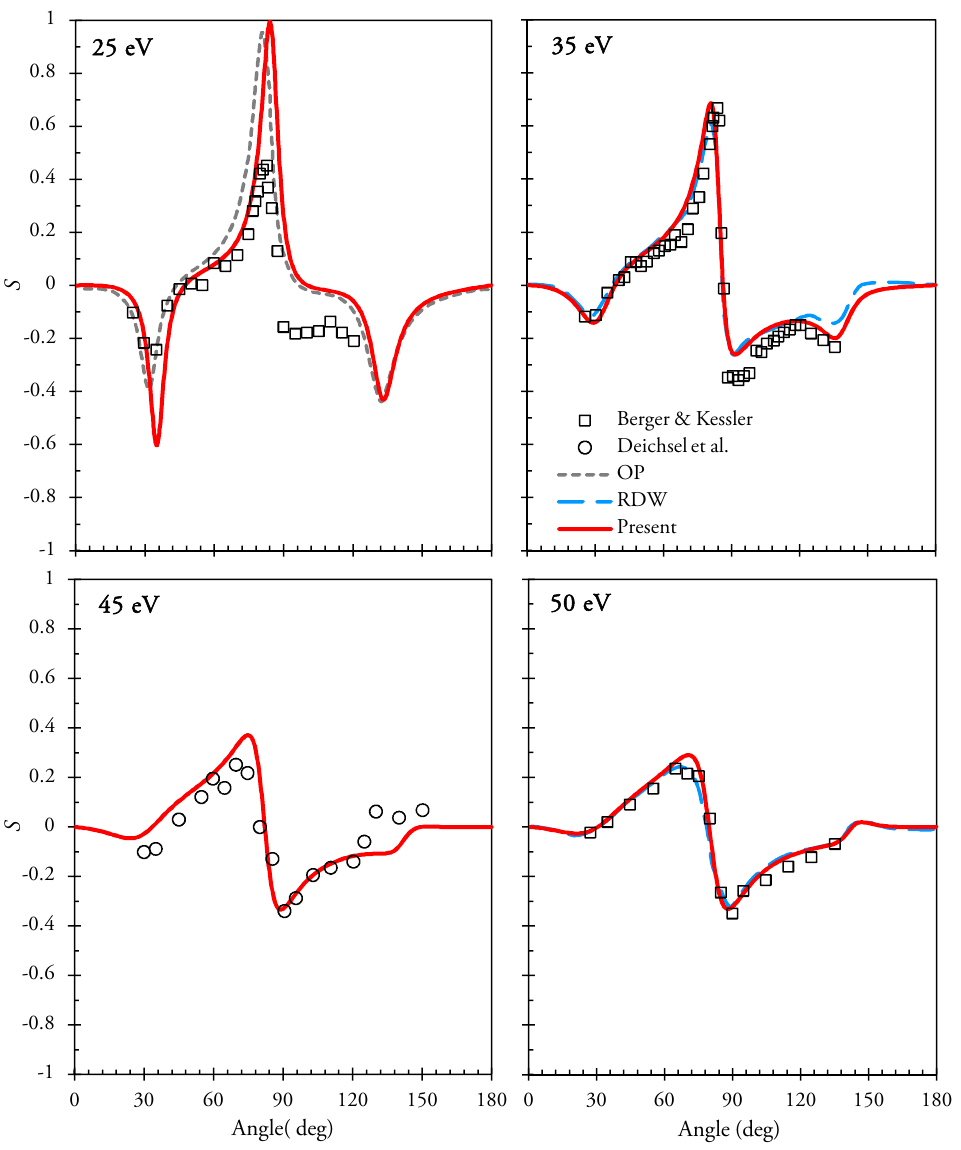}
\caption{\label{fig:hg6} Same as for figures~\ref{fig:hg2} and \ref{fig:hg5} but at 25, 35, 45 and 50 eV projectile energies. Additional experiment is Berger and Kessler~\cite{Berger1986}.}
\end{figure*}

\begin{figure*}
\includegraphics{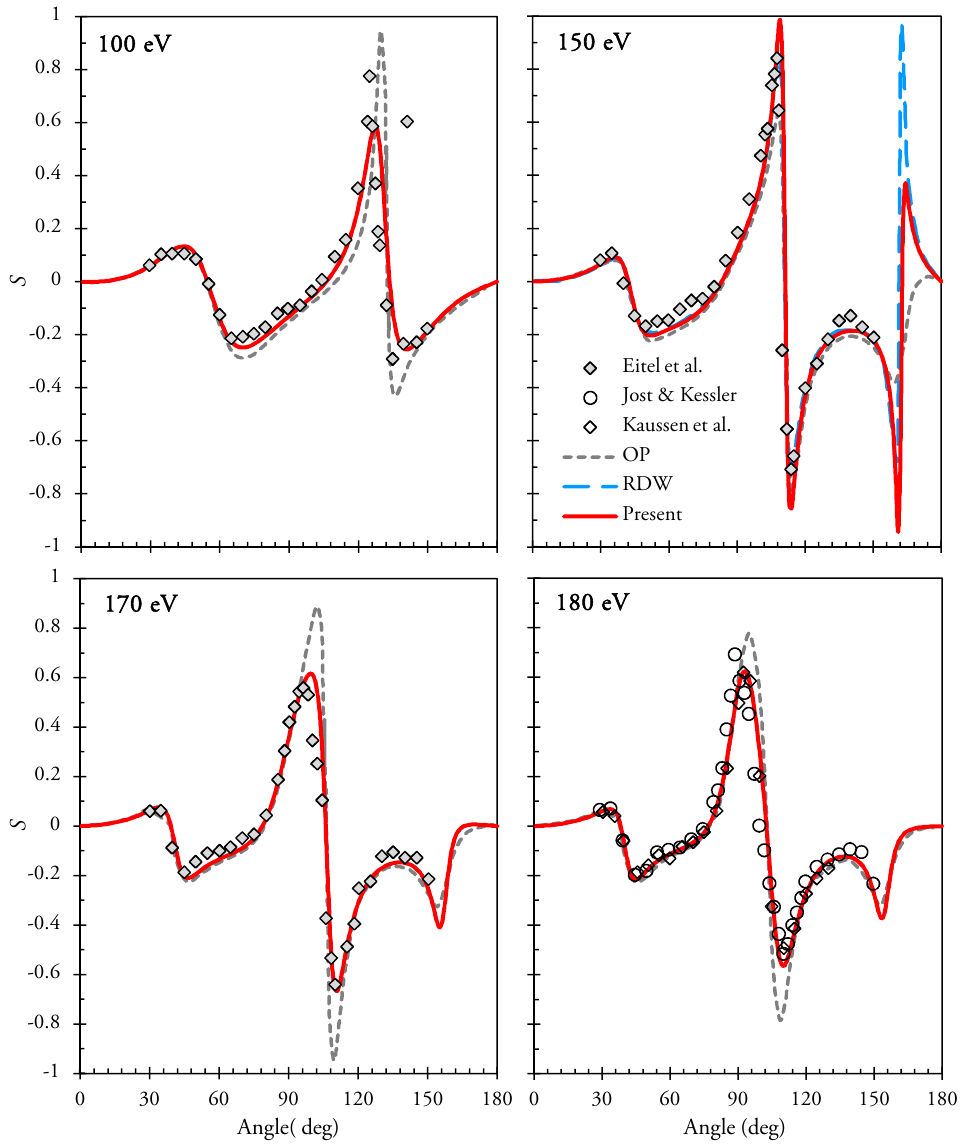}
\caption{\label{fig:hg7} Same as for figures~\ref{fig:hg6} but at 100, 150, 170 and 180 eV projectile energies. Additional experiments are Jost and Kessler~\cite{Jost1966} and Eitel et al.~\cite{Eitel1967}.}
\end{figure*}

\begin{figure*}
\includegraphics{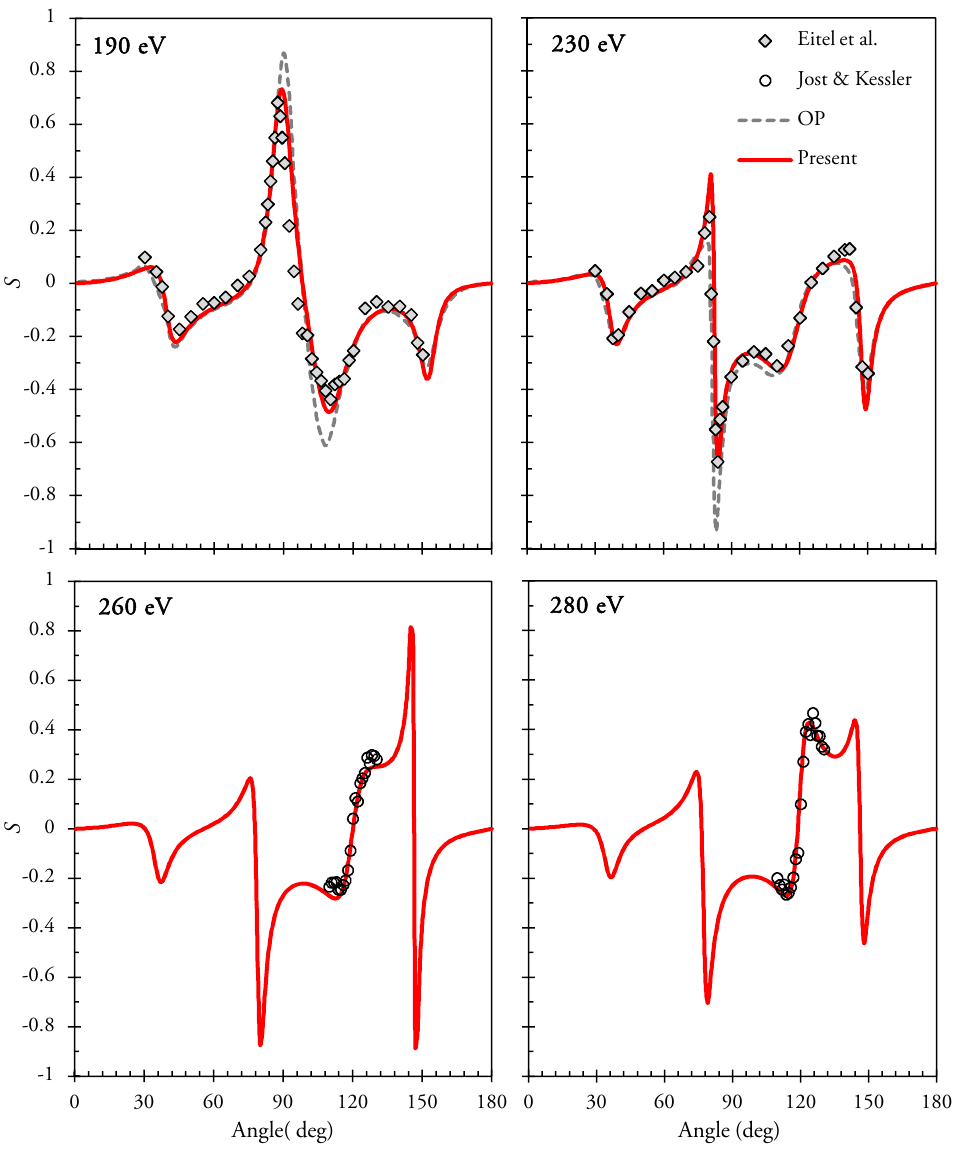}
\caption{\label{fig:hg8} Same as for figures~\ref{fig:hg7} but at 190, 230, 260 and 280 eV projectile energies.}
\end{figure*}

\begin{figure*}
\includegraphics{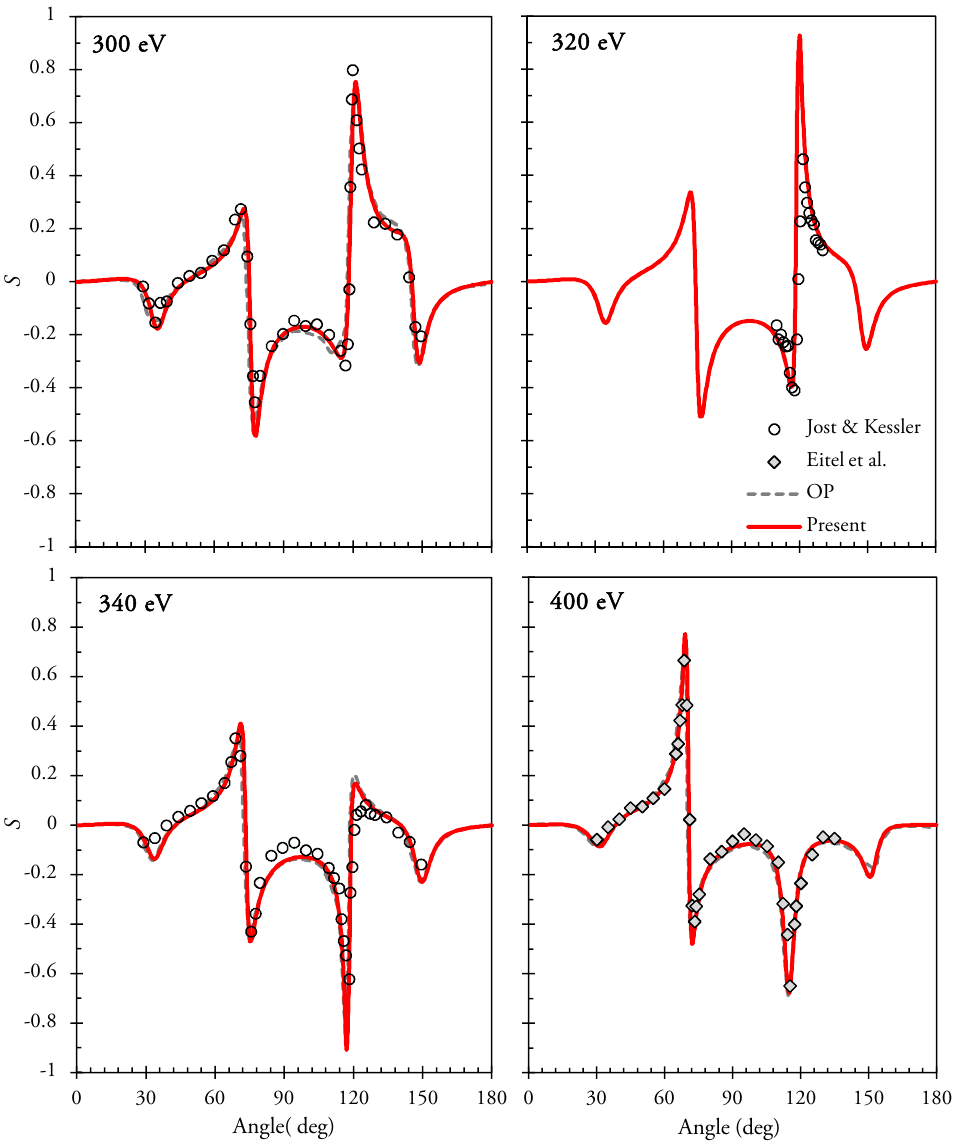}
\caption{\label{fig:hg9} Same as for figures~\ref{fig:hg8} but at 300, 320, 340 and 400 eV projectile energies.}
\end{figure*}

\begin{figure*}
\includegraphics{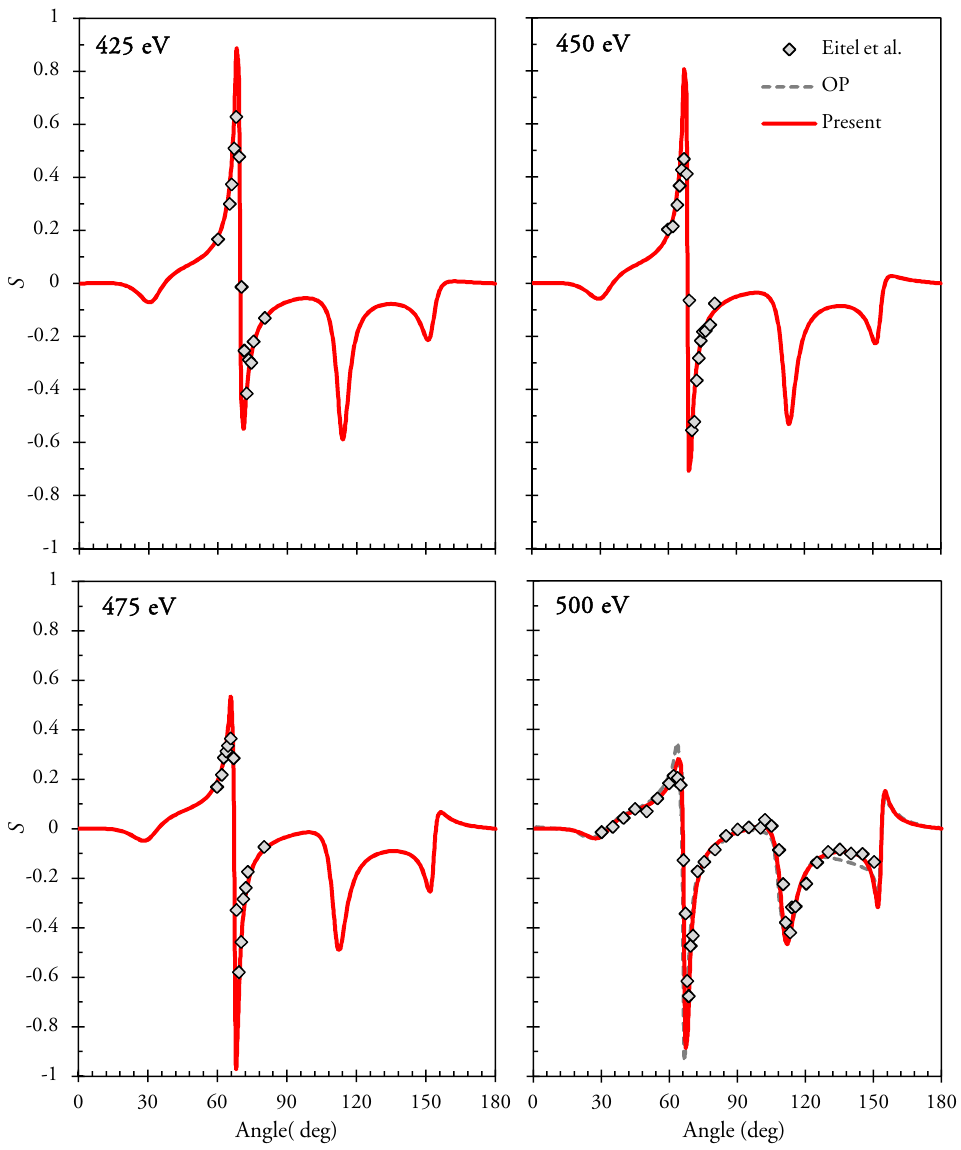}
\caption{\label{fig:hg10} Same as for figures~\ref{fig:hg9} but at 425, 450, 475 and 500 eV projectile energies.}
\end{figure*}

\begin{figure*}
\includegraphics{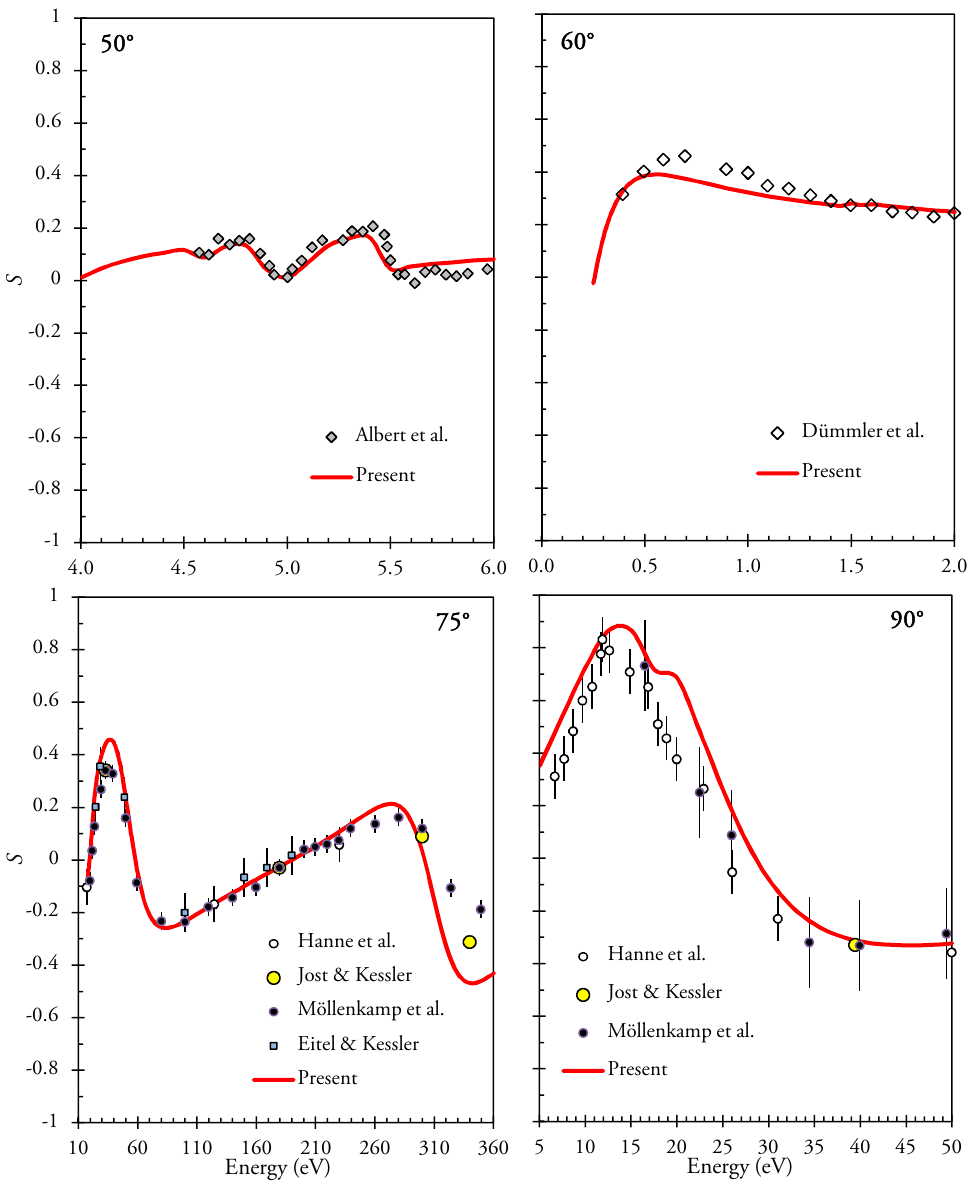}
\caption{\label{fig:hg11} Sherman function for electron scattering from mercury at $50^\circ$, $60^\circ$, $75^\circ$ and $90^\circ$ scattering angles. The legend in the figure describes markers for Present work; Experiment: Albert et al.~\cite{Albert1977}, D\"{u}mmler et al.~\cite{Dummler1992}, Hanne et al.~\cite{Hanne1980}, Jost and Kessler~\cite{Jost1966}, M\"{o}lenkamp et al.~\cite{Mollenkamp1984} and Eitel and Kessler~\cite{Eitel1971}}
\end{figure*}

In Figures~\ref{fig:hg1} -- \ref{fig:hg11}, we present a detailed comparison of our Sherman functions for mercury, from 1 eV through 500 eV, with experimental and select theoretical data. The theoretical works in the comparisons include the 66-state RCCC calculations of Bostock et al.~\cite{Bostock2012}, the R-matrix calculations of Bartschat et al.~\cite{Bartschat1984}, the generalized density-functional calculations (GDF) of Haberland and Fritsche~\cite{Haberland1987}, the relativistic distorted wave method (RDW) of McEachran and Stauffer~\cite{McEachran1987}, and the optical potential (OP) calculations of Haque et al.~\cite{Haque2017}. 

In figure~\ref{fig:hg1}, our Sherman functions display a very good agreement with the measurements of D\"{u}mmler et al.~\cite{Dummler1992} at energies 1, 1.5 and 1.9 eV. In the same figure and at energy 2.4 eV, the RCCC and our Sherman functions only exhibit partial agreement with the experimental data of  D\"{u}weke et al.~\cite{Duweke1976}. At energies 3.5, 3.9, 6 and 7 eV, present Sherman functions agree with experiment quite well but mainly in the forward direction, as shown in figure~\ref{fig:hg2}. 

In figure~\ref{fig:hg3}, our values agree with the measurements of Hanne et al.~\cite{Hanne1980} at impact energies 8 and 10 eV while throughout the angular range disagreeing with the OP predictions at those energies. In the same figure, our data show a considerably good agreement with the experiment of Kaussen et al.~\cite{Kaussen1987} and RCCC values at 9 and 11 eV.

We observe an excellent agreement, in figure~\ref{fig:hg4}, with experiment at 12 eV. Very good agreement with experiment is also on display at 12.2 and 14 eV, while not quite reaching the measured values at the two minima of $S(\theta)$. The agreement with experiment at 15 eV is better in the forward direction while notable disagreement is observed with the values of OP calculations.

Our Sherman functions, in figure~\ref{fig:hg5}, show good agreement with the experimental results of Kaussen et al.~\cite{Kaussen1987} at 17 and 24 eV. However, considerable disagreement is observed with the measurements of Hanne et al.~\cite{Hanne1980} at 18 eV and Deichsel et al.~\cite{Deichsel1966} at 23 eV. There are no other theoretical data to compare with at these energies. 

For energies above 23 eV it is clear that theory predicts accurately the experimentally confirmed physics of electron polarization during elastic scattering from mercury. In figure~\ref{fig:hg6}, present results display good agreement with experimental data while slightly disagreeing with those of OP on the positions of the first minimum and the maximum at 25 eV impact energy.  The RDW results follow nicely our values.

Present Sherman functions, in figure~\ref{fig:hg7}, display excellent agreement with experimental data. Our values differ from those of RDW at 150 eV only in the backward direction whereas visible differences are on display between our Sherman functions and those of the OP calculations at 100, 170 and 180 eV in figure~\ref{fig:hg7}.

At higher impact energies, our results are in excellent agreement with the available experimental data as shown in figures~\ref{fig:hg8}--~\ref{fig:hg10}.  No other calculations exist in this energy range, other than an OP one at 500 eV.

In figure~\ref{fig:hg11}, we compare our Sherman function values at $50^\circ$ with measurements of Albert et al.~\cite{Albert1977}, at $60^\circ$ with those of D\"{u}mmler et al.~\cite{Dummler1992}, and at $75^\circ$ and $90^\circ$ with the experimental results of Hanne et al.~\cite{Hanne1980}, Jost and Kessler~\cite{Jost1966}, M\"{o}lenkamp et al.~\cite{Mollenkamp1984} and Eitel and Kessler~\cite{Eitel1971} for various energy intervals. Overall the agreement with experimental data is very good.

\begin{figure*}
\includegraphics{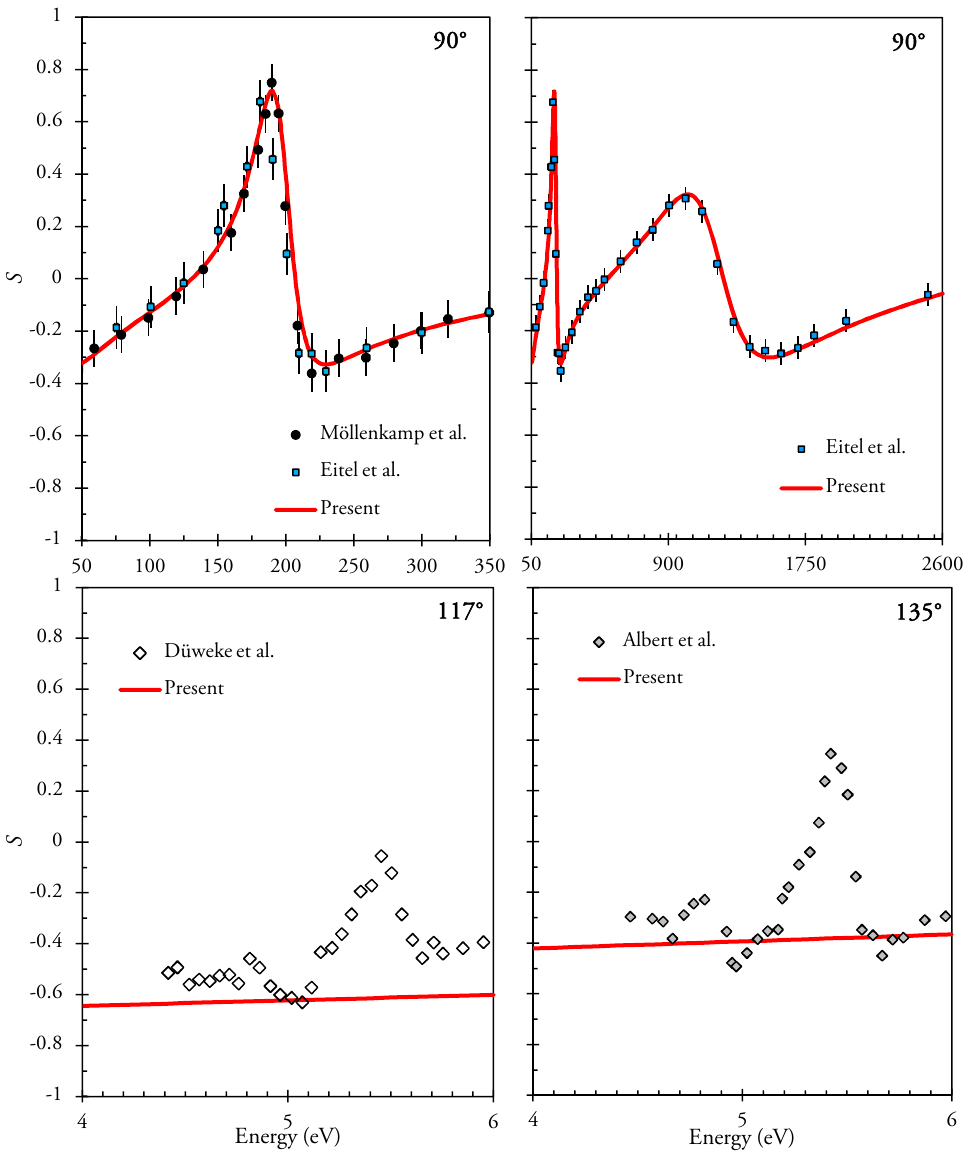}
\caption{\label{fig:hg12} Same as for figures~\ref{fig:hg11} but at $90^\circ$, $117^\circ$ and $135^\circ$ scattering angles. Additional experiment is D\"{u}weke et al.~\cite{Duweke1976}.}
\end{figure*}

Further comparisons with experimental data of M\"{o}lenkamp et al.~\cite{Mollenkamp1984} and Eitel et al.~\cite{Eitel1967} at $90^\circ$ are displayed in figure~\ref{fig:hg12} for energies up to 2600 eV exhibiting excellent agreement. In the same figure, our Sherman function values at $117^\circ$ and $135^\circ$ show a strong disagreement in the energy interval of 4 - 6 eV with measurements of D\"{u}weke et al.~\cite{Duweke1976} and Albert et al.~\cite{Albert1977}, respectively. This is not very surprising as our Sherman functions did not display good agreements with experimental data in backward direction at those energies (see figure~\ref{fig:hg2}). 

\begin{figure*}
\includegraphics{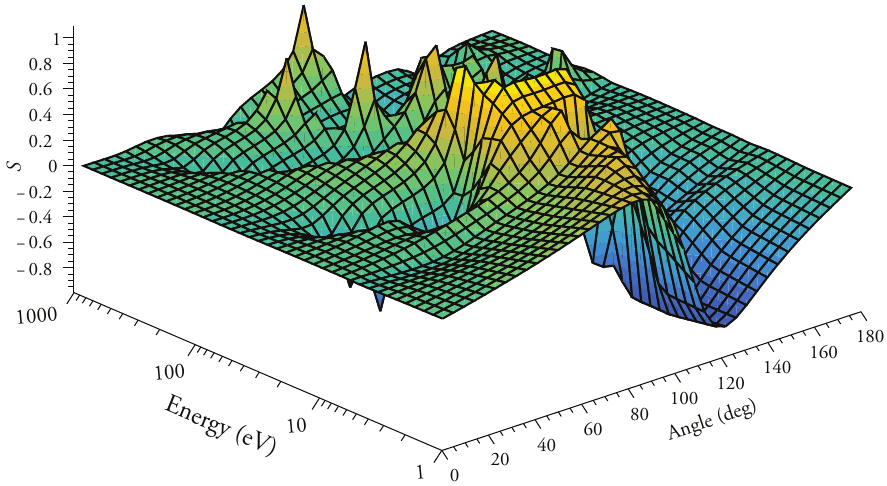}
\caption{\label{fig:hg13} A three-dimensional view of Sherman function for electron scattering from mercury.}
\end{figure*}

Finally, figure~\ref{fig:hg13} displays the 3D graph of Sherman function against impact energy and scattering angle for elastic $e$-Hg scattering to provide a global perspective. 

\section{Conclusions}
In this work we extended our previous work using a self-consistent Dirac-Slater calculation of the spin polarization of elastically scattered electrons from Zn, Cd, and Hg.  The central target potential was supplemented with a core-polarization term which contains the only adjustable parameter of this approach, i.e. the cut-off radius of the polarization potential.  We view the overall agreement with all the available experimental data and sophisticated calculations quite satisfactory.  The present comprehensive data aim to serve as a reliable reference point for future investigations.
  
\bibliography{zcmsa}

@article{Adibzadeh2004,
  title = {Elastic electron scattering from Ba and Sr},
  author = {Adibzadeh, Mehrdad and Theodosiou, Constantine E.},
  journal = {Phys. Rev. A},
  volume = {70},
  issue = {5},
  pages = {052704},
  numpages = {11},
  year = {2004},
  month = {Nov},
  publisher = {American Physical Society},
  doi = {10.1103/PhysRevA.70.052704},
  url = {https://link.aps.org/doi/10.1103/PhysRevA.70.052704}
}

@article{Adibzadeh2005,
title = {Elastic electron scattering from inert-gas atoms},
journal = {Atomic Data and Nuclear Data Tables},
volume = {91},
number = {1},
pages = {8-76},
year = {2005},
issn = {0092-640X},
doi = {https://doi.org/10.1016/j.adt.2005.07.004},
url = {https://www.sciencedirect.com/science/article/pii/S0092640X05000343},
author = {Mehrdad Adibzadeh and Constantine E. Theodosiou}
}

@Article{Adibzadeh2024,
AUTHOR = {Adibzadeh, Mehrdad and Theodosiou, Constantine E. and Harmon, Nicholas J.},
TITLE = {Elastic Electron Scattering from Be, Mg, and Ca},
JOURNAL = {Atoms},
VOLUME = {12},
YEAR = {2024},
NUMBER = {6},
ARTICLE-NUMBER = {33},
URL = {https://www.mdpi.com/2218-2004/12/6/33},
ISSN = {2218-2004},
DOI = {10.3390/atoms12060033}
}

@article{McEachran1992,
doi = {10.1088/0953-4075/25/7/022},
url = {https://dx.doi.org/10.1088/0953-4075/25/7/022},
year = {1992},
month = {apr},
publisher = {},
volume = {25},
number = {7},
pages = {1527},
author = {R P McEachran and  A D Stauffer},
title = {Spin polarization of electrons elastically scattered from cadmium and zinc},
journal = {Journal of Physics B: Atomic, Molecular and Optical Physics}
}

@article{Berrington2012,
  title = {Calculations of electron scattering from cadmium},
  author = {Berrington, Michael J. and Bostock, Christopher J. and Fursa, Dmitry V. and Bray, Igor and McEachran, R. P. and Stauffer, A. D.},
  journal = {Phys. Rev. A},
  volume = {85},
  issue = {4},
  pages = {042708},
  numpages = {21},
  year = {2012},
  month = {Apr},
  publisher = {American Physical Society},
  doi = {10.1103/PhysRevA.85.042708},
  url = {https://link.aps.org/doi/10.1103/PhysRevA.85.042708}
}

@article{Nahar1991,
  title = {Cross sections and spin polarizations for ${\mathit{e}}^{\ifmmode\pm\else\textpm\fi{}}$ scattering from cadmium},
  author = {Nahar, Sultana N.},
  journal = {Phys. Rev. A},
  volume = {43},
  issue = {5},
  pages = {2223--2236},
  numpages = {0},
  year = {1991},
  month = {Mar},
  publisher = {American Physical Society},
  doi = {10.1103/PhysRevA.43.2223},
  url = {https://link.aps.org/doi/10.1103/PhysRevA.43.2223}
}

@article{Szmytkowski1994,
doi = {10.1088/0953-4075/27/3/019},
url = {https://dx.doi.org/10.1088/0953-4075/27/3/019},
year = {1994},
month = {feb},
publisher = {},
volume = {27},
number = {3},
pages = {555},
author = {R Szmytkowski and  J E Sienkiewicz},
title = {Spin polarization of slow electrons elastically scattered from mercury, cadmium and zinc atoms},
journal = {Journal of Physics B: Atomic, Molecular and Optical Physics}
}

@article{Bartsch1992,
doi = {10.1088/0953-4075/25/7/021},
url = {https://dx.doi.org/10.1088/0953-4075/25/7/021},
year = {1992},
month = {apr},
publisher = {},
volume = {25},
number = {7},
pages = {1511},
author = {M Bartsch and  H Geesmann and  G F Hanne and  J Kessler},
title = {Asymmetric scattering of polarized electrons from atoms with closed and open shells},
journal = {Journal of Physics B: Atomic, Molecular and Optical Physics}
}

@article{Kumar1994,
  title = {Spin-polarization parameters and cross sections for electron scattering from zinc and lead atoms},
  author = {Kumar, Pradeep and Jain, Arvind Kumar and Tripathi, A. N. and Nahar, Sultana N.},
  journal = {Phys. Rev. A},
  volume = {49},
  issue = {2},
  pages = {899--907},
  numpages = {0},
  year = {1994},
  month = {Feb},
  publisher = {American Physical Society},
  doi = {10.1103/PhysRevA.49.899},
  url = {https://link.aps.org/doi/10.1103/PhysRevA.49.899}
}

@article{Dummler1992,
doi = {10.1088/0953-4075/25/20/022},
url = {https://dx.doi.org/10.1088/0953-4075/25/20/022},
year = {1992},
month = {oct},
publisher = {},
volume = {25},
number = {20},
pages = {4281},
author = {M D\"{u}mmler and  M Bartsch and  H Geesmann and  G F Hanne and  J Kessler},
title = {Low-energy data of the Sherman function of Hg, Tl and Pb},
journal = {Journal of Physics B: Atomic, Molecular and Optical Physics}
}

@article{Haberland1987,
doi = {10.1088/0022-3700/20/1/016},
url = {https://dx.doi.org/10.1088/0022-3700/20/1/016},
year = {1987},
month = {jan},
publisher = {},
volume = {20},
number = {1},
pages = {121},
author = {R Haberland and  L Fritsche},
title = {On the elastic scattering of low-energy electrons by Hg, Tl, Pb and Bi atoms},
journal = {Journal of Physics B: Atomic and Molecular Physics}
}

@article{Kaussen1987,
doi = {10.1088/0022-3700/20/1/018},
url = {https://dx.doi.org/10.1088/0022-3700/20/1/018},
year = {1987},
month = {jan},
publisher = {},
volume = {20},
number = {1},
pages = {151},
author = {F Kaussen and  H Geesmann and  G F Hanne and  J Kessler},
title = {Study of spin polarisation in elastic scattering of electrons from Hg, Tl, Pb and Bi atoms},
journal = {Journal of Physics B: Atomic and Molecular Physics}
}

@article{McEachran1987,
doi = {10.1088/0022-3700/20/20/027},
url = {https://dx.doi.org/10.1088/0022-3700/20/20/027},
year = {1987},
month = {oct},
publisher = {},
volume = {20},
number = {20},
pages = {5517},
author = {R P McEachran and  A D Stauffer},
title = {Spin polarisation of electrons elastically scattered from mercury},
journal = {Journal of Physics B: Atomic and Molecular Physics}
}

@article{Sienkiewicz1990,
doi = {10.1088/0953-4075/23/11/020},
url = {https://dx.doi.org/10.1088/0953-4075/23/11/020},
year = {1990},
month = {jun},
publisher = {},
volume = {23},
number = {11},
pages = {1869},
author = {J E Sienkiewicz},
title = {Spin polarisation and differential cross sections in elastic low-energy scattering of electrons from mercury},
journal = {Journal of Physics B: Atomic, Molecular and Optical Physics}
}

@article{Fritsche1992,
doi = {10.1088/0953-4075/25/20/023},
url = {https://dx.doi.org/10.1088/0953-4075/25/20/023},
year = {1992},
month = {oct},
publisher = {},
volume = {25},
number = {20},
pages = {4287},
author = {L Fritsche and  C Kroner and  T Reinert},
title = {A consistent relativistic theory of electron scattering by atoms},
journal = {Journal of Physics B: Atomic, Molecular and Optical Physics}
}

@incollection{Haque2021,
title = {Chapter One - Elastic scattering of e± by Cd, Hg, and Pb atoms at 1 eV ≤ Ei ≤ 1 GeV},
editor = {Erkki J. Brändas},
series = {Advances in Quantum Chemistry},
publisher = {Academic Press},
volume = {84},
pages = {1-72},
year = {2021},
issn = {0065-3276},
doi = {https://doi.org/10.1016/bs.aiq.2020.03.002},
url = {https://www.sciencedirect.com/science/article/pii/S0065327620300046},
author = {Mohammad M. Haque and Abul K.F. Haque and Mohammad Alfaz Uddin and Malik Maaza and Mohammad Atiqur R. Patoary and Arun K. Basak and Bidhan C. Saha}
}

@article{Kolb1982,
  title = {Electric and magnetic susceptibilities and shielding factors for closed-shell atoms and ions of high nuclear charge},
  author = {Kolb, Dietmar and Johnson, W. R. and Shorer, Philip},
  journal = {Phys. Rev. A},
  volume = {26},
  issue = {1},
  pages = {19--31},
  numpages = {0},
  year = {1982},
  month = {Jul},
  publisher = {American Physical Society},
  doi = {10.1103/PhysRevA.26.19},
  url = {https://link.aps.org/doi/10.1103/PhysRevA.26.19}
}

@article{Berger1986,
doi = {10.1088/0022-3700/19/21/018},
url = {https://dx.doi.org/10.1088/0022-3700/19/21/018},
year = {1986},
month = {nov},
publisher = {},
volume = {19},
number = {21},
pages = {3539},
author = {O Berger and J Kessler},
title = {Elastic scattering of polarised electrons from mercury and xenon},
journal = {Journal of Physics B: Atomic and Molecular Physics}
}

@article{Duweke1976,
doi = {10.1088/0022-3700/9/11/017},
url = {https://dx.doi.org/10.1088/0022-3700/9/11/017},
year = {1976},
month = {aug},
publisher = {},
volume = {9},
number = {11},
pages = {1915},
author = {M D\"{u}weke and N Kirchner and E Reichert and S Sch\"{o}n},
title = {Spin polarization in low-energy electron scattering from mercury},
journal = {Journal of Physics B: Atomic and Molecular Physics}
}

@article{Bostock2012,
  title = {Relativistic convergent close-coupling method calculation of the spin polarization of electrons scattered elastically from zinc and mercury},
  author = {Bostock, Christopher J. and Fursa, Dmitry V. and Bray, Igor},
  journal = {Phys. Rev. A},
  volume = {85},
  issue = {6},
  pages = {062707},
  numpages = {6},
  year = {2012},
  month = {Jun},
  publisher = {American Physical Society},
  doi = {10.1103/PhysRevA.85.062707},
  url = {https://link.aps.org/doi/10.1103/PhysRevA.85.062707}
}

@article{Kelemen2012,
doi = {10.1088/0953-4075/45/18/185202},
url = {https://dx.doi.org/10.1088/0953-4075/45/18/185202},
year = {2012},
month = {sep},
publisher = {IOP Publishing},
volume = {45},
number = {18},
pages = {185202},
author = {Kelemen, V I and Remeta, E Yu},
title = {Critical minima and spin polarization in the elastic electron scattering by the mercury atoms},
journal = {Journal of Physics B: Atomic, Molecular and Optical Physics}
}

@article{Deichsel1966,
  title={Elektronenpolarisation im Energiebereich unterhalb 50 eV durch Streuung an freien Hg-Atomen},
  author={Deichsel, H and Reichert, E and Steidl, H},
  journal={Zeitschrift f{\"u}r Physik},
  volume={189},
  number={2},
  pages={212--216},
  year={1966},
  publisher={Springer}
}

@article{Jost1966,
  title={Zur Polarisation langsamer Elektronen durch Streuung an Quecksilber zwischen 180 und 1700 eV},
  author={Jost, Klaus and Kessler, Joachim},
  journal={Zeitschrift f{\"u}r Physik},
  volume={195},
  pages={1--12},
  year={1966},
  publisher={Springer}
}

@article{Hanne1980,
doi = {10.1088/0022-3700/13/12/009},
url = {https://dx.doi.org/10.1088/0022-3700/13/12/009},
year = {1980},
month = {jun},
publisher = {},
volume = {13},
number = {12},
pages = {L395},
author = {G F Hanne and K J Kollath and W Wubker},
title = {Electron spin polarisation resulting from elastic scattering of electrons by mercury atoms between 8 and 18 eV collision energy},
journal = {Journal of Physics B: Atomic and Molecular Physics}
}

@article{Eitel1967,
  title = {Polarization of Slow Electrons by Hg and Range of Applicability of the Relativistic Hartree Potential},
  author = {Eitel, W. and Jost, K. and Kessler, J.},
  journal = {Phys. Rev.},
  volume = {159},
  issue = {1},
  pages = {47--49},
  numpages = {0},
  year = {1967},
  month = {Jul},
  publisher = {American Physical Society},
  doi = {10.1103/PhysRev.159.47},
  url = {https://link.aps.org/doi/10.1103/PhysRev.159.47}
}

@article{Furness1973,
doi = {10.1088/0022-3700/6/11/021},
url = {https://dx.doi.org/10.1088/0022-3700/6/11/021},
year = {1973},
month = {nov},
publisher = {},
volume = {6},
number = {11},
pages = {2280},
author = {J B Furness and  I E McCarthy},
title = {Semiphenomenological optical model for electron scattering on atoms},
journal = {Journal of Physics B: Atomic and Molecular Physics}
}

@article{Fink1969,
title = {Theoretical electron scattering amplitudes and spin polarizations: Selected targets, electron energies 100 to 1500 eV},
journal = {Atomic Data and Nuclear Data Tables},
volume = {1},
pages = {385-456},
year = {1969},
issn = {0092-640X},
doi = {https://doi.org/10.1016/S0092-640X(69)80029-X},
url = {https://www.sciencedirect.com/science/article/pii/S0092640X6980029X},
author = {Manfred Fink and Albert C. Yates}
}

@article{Berger1981,
  title = {"Triple" Scattering Experiment for Obtaining the Maximum Possible Information on Elastic Electron Scattering from Mercury},
  author = {Berger, O. and Kessler, J. and Kollath, K. J. and M\"ollenkamp, R. and W\"ubker, W.},
  journal = {Phys. Rev. Lett.},
  volume = {46},
  issue = {12},
  pages = {768--770},
  numpages = {0},
  year = {1981},
  month = {Mar},
  publisher = {American Physical Society},
  doi = {10.1103/PhysRevLett.46.768},
  url = {https://link.aps.org/doi/10.1103/PhysRevLett.46.768}
}

@article{Hanne1972,
  title={Spinpolarisation und Winkelverteilung der Elektronen nach Anregung verschiedener Zust{\"a}nde des Hg-Atoms},
  author={Hanne, GF and Jost, K and Kessler, J},
  journal={Zeitschrift f{\"u}r Physik},
  volume={252},
  number={2},
  pages={141--146},
  year={1972},
  publisher={Springer}
}

@article{Bunyan1963,
doi = {10.1088/0370-1328/81/5/304},
url = {https://doi.org/10.1088/0370-1328/81/5/304},
year = {1963},
month = {may},
publisher = {},
volume = {81},
number = {5},
pages = {816},
author = {P J Bunyan},
title = {The Polarization by Mercury of 1 to 2 keV Electrons},
journal = {Proceedings of the Physical Society}
}

@article{Bunyan1965,
doi = {10.1088/0370-1328/85/3/306},
url = {https://doi.org/10.1088/0370-1328/85/3/306},
year = {1965},
month = {mar},
publisher = {},
volume = {85},
number = {3},
pages = {455},
author = {P J Bunyan and J L Schonfelder},
title = {Polarization by mercury of 100 to 2000 eV electrons},
journal = {Proceedings of the Physical Society}
}

@article{Bartschat1984,
doi = {10.1088/0022-3700/17/18/016},
url = {https://doi.org/10.1088/0022-3700/17/18/016},
year = {1984},
month = {sep},
publisher = {},
volume = {17},
number = {18},
pages = {3797},
author = {K Bartschat and K Blum and P G Burke and G F Hanne and N S Scott},
title = {The fine-structure effect in the low-energy scattering of electrons on Hg and Tl atoms},
journal = {Journal of Physics B: Atomic and Molecular Physics}
}

@article{Deichsel1961,
  title={Herstellung und Nachweis polarisierter Elektronenstrahlen durch zweimalige Streuung von Gl{\"u}helektronen kleiner Energie (1--2 keV) an Hg-Atomstrahlen},
  author={Deichsel, Heinrich},
  journal={Zeitschrift f{\"u}r Physik},
  volume={164},
  number={2},
  pages={156--165},
  year={1961},
  publisher={Springer}
}

@article{Mollenkamp1984,
doi = {10.1088/0022-3700/17/6/022},
url = {https://doi.org/10.1088/0022-3700/17/6/022},
year = {1984},
month = {mar},
publisher = {},
volume = {17},
number = {6},
pages = {1107},
author = {R M\"{o}llenkamp and W W\"{u}bker and O Berger and K Jost and J Kessler},
title = {Elastic scattering of polarised electrons from mercury and xenon to obtain the complete information on the scattering process},
journal = {Journal of Physics B: Atomic and Molecular Physics}
}

@article{Albert1977,
doi = {10.1088/0022-3700/10/18/029},
url = {https://doi.org/10.1088/0022-3700/10/18/029},
year = {1977},
month = {dec},
publisher = {},
volume = {10},
number = {18},
pages = {3733},
author = {K Albert and C Christian and T Heindorff and E Reichert and S Sch\"{o}n},
title = {Electron resonance scattering from mercury},
journal = {Journal of Physics B: Atomic and Molecular Physics}
}

@article{Sherman1956,
  title = {Coulomb Scattering of Relativistic Electrons by Point Nuclei},
  author = {Sherman, Noah},
  journal = {Phys. Rev.},
  volume = {103},
  issue = {6},
  pages = {1601--1607},
  numpages = {0},
  year = {1956},
  month = {Sep},
  publisher = {American Physical Society},
  doi = {10.1103/PhysRev.103.1601},
  url = {https://link.aps.org/doi/10.1103/PhysRev.103.1601}
  }

@article{Kessler1969,
  title = {Electron Spin Polarization by Low-Energy Scattering from Unpolarized Targets},
  author = {Kessler, J.},
  journal = {Rev. Mod. Phys.},
  volume = {41},
  issue = {1},
  pages = {3--25},
  numpages = {0},
  year = {1969},
  month = {Jan},
  publisher = {American Physical Society},
  doi = {10.1103/RevModPhys.41.3},
  url = {https://link.aps.org/doi/10.1103/RevModPhys.41.3}
}

@article{Eitel1971,
  title={Spinpolarisation und differentielle Wirkungsquerschnitte bei der inelastischen Elektron-Atom-Streuung},
  author={Eitel, Wilhelm and Kessler, J},
  journal={Zeitschrift f{\"u}r Physik A Hadrons and nuclei},
  volume={241},
  number={4},
  pages={355--368},
  year={1971},
  publisher={Springer}
}

@article{Haque2017,
doi = {10.1088/2399-6528/aa8bf8},
url = {https://doi.org/10.1088/2399-6528/aa8bf8},
year = {2017},
month = {nov},
publisher = {IOP Publishing},
volume = {1},
number = {3},
pages = {035014},
author = {Haque, A K F and Haque, M M and Bhattacharjee, Prajna P and Uddin, M Alfaz and Patoary, M Atiqur R and Hossain, M Ismail and Basak, A K and Mahbub, M Selim and Maaza, M and Saha, B C},
title = {Relativistic calculations for spin-polarization of elastic electron—mercury scattering},
journal = {Journal of Physics Communications}
}

@article{Bostock2011,
  title = {Relativistic and Close-Coupling Effects in the Spin Polarization of Low-Energy Electrons Scattered Elastically from Cadmium},
  author = {Bostock, Christopher J. and Berrington, Michael J. and Fursa, Dmitry V. and Bray, Igor},
  journal = {Phys. Rev. Lett.},
  volume = {107},
  issue = {9},
  pages = {093202},
  numpages = {4},
  year = {2011},
  month = {Aug},
  publisher = {American Physical Society},
  doi = {10.1103/PhysRevLett.107.093202},
  url = {https://link.aps.org/doi/10.1103/PhysRevLett.107.093202}
}

@article{Salvat1995,
title = {Accurate numerical solution of the radial Schrödinger and Dirac wave equations},
journal = {Computer Physics Communications},
volume = {90},
number = {1},
pages = {151-168},
year = {1995},
issn = {0010-4655},
doi = {https://doi.org/10.1016/0010-4655(95)00039-I},
url = {https://www.sciencedirect.com/science/article/pii/001046559500039I},
author = {F. Salvat and J.M. Fernández-Varea and W. Williamson}
}

\end{document}